# Hiding Dust around ϵ Eridani

Schuyler Grace Wolff[1], András Gáspár[1], George H. Rieke[1], Nicholas Ballering[2], and Marie Ygouf[3]
[1] Steward Observatory and the Department of Astronomy, The University of Arizona, 933 N Cherry Avenue, Tucson, AZ 85719, USA; sgwolff@arizona.edu
[2] Department of Astronomy, University of Virginia, 530 McCormick Road Charlottesville, VA 22904, USA
[3] Jet Propulsion Laboratory, California Institute of Technology, Pasadena, CA, USA

## Abstract

With a Jupiter-like exoplanet and a debris disk with both asteroid and Kuiper Belt analogs, $\epsilon$ Eridani has a fascinating resemblance to our expectations for a young solar system. We present a deep Hubble Space Telescope/Space Telescope Imaging Spectrograph coronographic data set using eight orbit visits and the point-spread function calibrator $\delta$ Eridani. While we were unable to detect the debris disk, we place stringent constraints on the scattered light surface brightness of $\sim 4\,\mu$Jy arcsec$^{-2}$. We combine this scattered light detection limit with a reanalysis of archival near- and mid-infrared observations and a dynamical model of the full planetary system to refine our model of the $\epsilon$ Eridani debris disk components. Radiative transfer modeling suggests an asteroid belt analog inside of 3 au, an intermediate disk component in the 6–37 au region, and a Kuiper Belt analog colocated with the narrow belt observed in the millimeter (69 au). Modeling also suggests a large minimum grain size requiring either very porous grains or a suppression of small grain production, and a radially stratified particle size distribution. The inner disk regions require a steep power-law slope ($s^{-3.8}$ where $s$ is the grain size) weighted toward smaller grains and the outer disk prefers a shallower slope ($s^{-3.4}$) with a minimum particle size of $>2\,\mu$m. These conclusions will be enhanced by upcoming coronagraphic observations of the system with the James Webb Space Telescope, which will pinpoint the radial location of the dust belts and further diagnose the dust particle properties.

*Unified Astronomy Thesaurus concepts:* Coronagraphic imaging (313); Circumstellar dust (236); Debris disks (363)

## 1. Introduction

The best-studied planetary system, without a doubt, is our very own. Its structure, with rocky terrestrial planets on inner orbits, gaseous and ice giant planets on outer orbits, and belts of debris associated with the freezing points of water (the asteroid belt) and other compounds (the Kuiper Belt), was once considered to be the archetypal layout. Since their formation near the freezing points of various molecules, the locations of the belts have been heavily influenced by the gravitational perturbations of nearby planets, with Jupiter sculpting and constraining the orbits of asteroids in the inner system and Neptune shepherding the inner edge of the Kuiper Belt. If we did not know about the planets but could image these two debris belts, we could infer the presence of Jupiter and Neptune from their key roles in shaping the morphology of these belts. This is an important observation since most exoplanets appear to be of Neptune or lower mass (e.g., Malhotra 2015) and hence well out of reach for direct detection; yet, the positions and morphology of debris belts can indicate their presence and constrain their orbits. Large-scale infrared surveys have inferred similar architecture for ∼30% of known circumstellar debris systems (e.g., Ballering et al. 2013; Chen et al. 2014), i.e., with a warm debris belt analogous to our asteroid belt and a cold one analogous to the Kuiper Belt. Nearly all of the remaining systems can be well described by a single belt model with temperatures indicating that they are either asteroidal-like or Kuiper-like (e.g., Morales et al. 2011; Ballering et al. 2017).

$\epsilon$ Eridani ($\epsilon$ Eri) is one of the most intriguing nearby stellar systems. At a distance of only 3.2 pc, it hosts the nearest prominent debris disk. Furthermore, it has at least one planet. $\epsilon$ Eri b was first discovered via a radial velocity detection by Hatzes et al. (2000), though the detection was complicated by stellar activity from this relatively young (200–800 Myr; Mamajek & Hillenbrand 2008) K2 spectral-type star. Recently, Mawet et al. (2019) combined new high-fidelity RV data with direct imaging constraints to arrive at a planetary mass of $M \sin i \sim 0.8\, M_{\rm Jup}$ with a separation of $3.48 \pm 0.02$ au (1″.1), a result that has been updated to $0.66^{+0.12}_{-0.09}\, M_{\rm Jup}$ (Llop-Sayson et al. 2021).

The debris disk was first detected via infrared excess seen with IRAS (Aumann 1985). Resolved imaging with the Atacama Large Millimeter/submillimeter Array (ALMA) shows a remarkably narrow outer ring centered at 69 au with a width of $12 \pm 1$ au (Booth et al. 2017). Close-in debris, i.e., within a radius of 1.5 au, has been seen through nulling interferometry at 10 $\mu$m (Ertel et al. 2020), which we can roughly describe as an inner disk or asteroid belt analog. These features are typical of debris systems, but atypically there appears to be a third component, i.e., an intermediate belt outside the region probed with LBTI, i.e., between the innermost asteroid belt analog and the outer Kuiper Belt analog (Backman et al. 2009; Greaves et al. 2014), but with a poorly constrained structure (Su et al. 2017). An alternative proposal is that this inner warm dust is being produced in the outer ring and pulled in via drag forces, resulting in a population of smaller grains with a flatter size distribution (Reidemeister et al. 2011).

So far, the debris system has not been detected in the visible, i.e., in scattered light. The structure of the inner parts of the system might be revealed at these wavelengths, as has been the case for many other debris systems probed down to the few arcsecond level (e.g., Schneider et al. 2014). Proffitt et al. (2004) imaged the outer disk with Hubble Space Telescope (HST) in a program designed to place the star itself 5″ off the field of view, and without using coronagraphy. Their reduction using roll deconvolution will result in self-cancellation of the ring signal, since the ring is nearly face-on, so the results are difficult to interpret. Their direct point-spread function (PSF) subtraction results show a rapidly degrading residual surface brightness approaching the star, with a value of ∼3 $\mu$Jy arcsec$^{-2}$ at 10″ (using the calibration from Proffitt 2005) and rising very rapidly inward. In addition, at a separation of 10″, less than one-third of a ring would fall on the detector array. Since the intermediate belt is thought to lie within 10″ = 32 au, these measurements are not useful for exploring its structure.

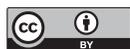

The structure of the debris system suggests that there may be an additional planet in the outer system. The outer debris ring at 69 au has sharp edges; Booth et al. (2017) suggested they may be set by resonances with an unseen planet. They show that its orbit would be expected to have a semimajor axis of <55 au with the highest likelihood at 48 au. This situation highlights that the $\epsilon$ Eri system can be resolved at a physical level that can probe sensitively for additional unseen planets through their gravitational effects on its complex debris system.

To advance our understanding, we have obtained HST/Space Telescope Imaging Spectrograph (STIS) coronagraphic observations of this debris system. After reference star subtraction, our results achieve a useful inner working angle of ∼2″ = 6 au. Inside of 10″, they provide a much more sensitive probe than is possible with the images of Proffitt et al. (2004). We combine the deep image sensitivities with constraints from dynamical modeling and photometry from the literature to refine the debris disk model in light of these new measurements. A description of both the STIS observations and a reanalysis of the near- and mid-IR properties of the system is provided in Section 2. Section 3 demonstrates several techniques to remove the stellar PSF via reference image subtraction in our coronographic observations. Data analysis and modeling are presented in Section 4. This includes dynamical modeling of the system in combination with radiative transfer modeling of the debris disk in a Monte Carlo analysis using spectral information and broadband photometry. A discussion of these results and a revised debris disk model are presented in Section 5.

## 2. New Observations and Additional Measurements

### 2.1. STIS Coronagraphy

The wide bandpass and 0″.05 pixel$^{-1}$ plate scale make STIS coronagraphic observations extremely sensitive to low surface brightness disk material. Our Cycle 27 HST program (PID GO15906) observed $\epsilon$ Eri and its PSF calibrator target ($\delta$ Eridani) for eight orbit visits using the instrument's 50CCD detector with the coronagraphic mask. The program was split into two groups of four visits, with three orbits dedicated for target and one for PSF observations within each group (with the PSF observed in the third visit). The target visits were designed to be offset by a 30° rotational dither, thereby providing high spatial sampling and full roll coverage. Due to limited guide star availability, there were necessary offsets from this optimal rotational dither pattern. In the first group of four visits, we placed $\epsilon$ Eri and its PSF target behind the WEDGEA1.8 and WEDGEB1.8 occulting positions on the coronagraphic masks. Based on the data collected during the first epoch, we re-planned our second epoch of observations to use the slightly wider (and therefore larger inner working angle) WEDGEA2.5 and WEDGEB2.5 occulting positions. The re-planning was initiated by a tentative detection of a ring around $\epsilon$ Eri following the reduction of only the first epoch of data. These wider positions enabled a 2.3× longer integration without saturation. The "detection," unfortunately, turned out to be likely an imaging artifact. In Table 1, we summarize the data taken within our program.

### 2.2. Photometry

In addition to the STIS coronagraphic data set, we also use a spectral energy distribution (SED) compiled from the literature to investigate the debris disk environment around $\epsilon$ Eri. The full SED information is provided in Appendix A for the overall, unresolved system. In the following sections, we aim to distinguish how the SEDs of the inner and intermediate disk regions contribute to the total. We first examine the near-IR excess (<8 $\mu$m), finding a lack of warm dust emission and then investigate the mid-IR disk features using archival Spitzer IRS information.

#### 2.2.1. Lack of Near-IR Excess

Here we focus on the photometric contributions from the inner region of the system. Unlike many stars that have small hot excesses in the near-IR, the emission there from $\epsilon$ Eri appears to be purely photospheric. There is no excess around $\epsilon$ Eri to within ∼0.5% of the photospheric emission at 2 $\mu$m within ∼1″ (Di Folco et al. 2007). We can extend this conclusion to the intermediate wavelengths using saturated IRAC photometry, which has been shown to have ≲1% errors (Su et al. 2022). We have taken the values relative to Sirius from Marengo et al. (2009). We have also determined an accurate value relative to Sirius for the $K_S$ band. This was generated using photometry from Glass (1974), Carter (1990), and Bouchet et al. (1991), transformed to the Two Micron All Sky Survey $K_S$ as described in Rieke et al. (2022), who demonstrated the accuracy of the transformed photometry (see Appendices A–C). We then generated models of Sirius and $\epsilon$ Eri from the BOSZ model set (Bohlin et al. 2017) and computed the ratio. Synthetic photometry was conducted on this ratio for all five bands ($K_S$ and the IRAC bands) and compared with the measured values. We made a small adjustment (<1%) in the overall level to minimize the residuals. The remaining scatter is 0.5% rms. That is, combined with the interferometric result for small apertures, we conclude that $\epsilon$ Eri has no infrared excess at wavelengths of 8 $\mu$m and shorter at the level of 1% of the stellar output.

#### 2.2.2. Mid-IR Observations of the Inner Debris System

The SED of the inner and intermediate debris disk components is shown in Figure 1, and photometry of this region is listed in Table 2. Spitzer IRS high-resolution spectroscopy using the Short High (SH: 10–20 $\mu$m) and Long High (LH: 19–37 $\mu$m) modules was first published for $\epsilon$ Eri in Backman et al. (2009) with data taken at two nod positions for each mode with total integration times of 72 s (SH) and 144 s (LH) per slit position. These data were reanalyzed in Su et al. (2017; see their Figure 6) wherein the extracted spectrum was taken from the CASSIS database (Lebouteiller et al. 2015),

Table 1
Observations within Our GO 15906 Program

| Visit | Target | Date | Apertures | Orientation (degrees) | Integration Time (s) | $N_{\text{Images}}$[a] |
|---|---|---|---|---|---|---|
| 1 | $\epsilon$ Eridani | 11-20-2019 | WEDGEA1.8,WEDGEB1.8 | −64.32 | 319.3 | 17,14 |
| 2 | $\epsilon$ Eridani | 11-20-2019 | WEDGEA1.8,WEDGEB1.8 | −41.32 | 319.3 | 17,14 |
| 3 | $\delta$ Eridani (PSF) | 11-20-2019 | WEDGEA1.8,WEDGEB1.8 | | 257.3 | 17,14 |
| 4 | $\epsilon$ Eridani | 11-20-2019 | WEDGEA1.8,WEDGEB1.8 | −18.32 | 319.3 | 17,14 |
| 5 | $\epsilon$ Eridani | 01-28-2021 | WEDGEA2.5,WEDGEB2.5 | 3.06 | 728.0 | 13,13 |
| 6 | $\epsilon$ Eridani | 01-28-2021 | WEDGEA2.5,WEDGEB2.5 | 33.06 | 728.0 | 13,13 |
| 7 | $\delta$ Eridani (PSF) | 01-28-2021 | WEDGEA2.5,WEDGEB2.5 | | 587.6 | 13,13 |
| 8 | $\epsilon$ Eridani | 01-28-2021 | WEDGEA2.5,WEDGEB2.5 | 47.42 | 728.0 | 13,13 |

**Note.**
[a] Number of images taken at each coronagraphic aperture.

Table 2
Photometry of the $\epsilon$ Eri Inner Debris Disk

| Wavelength ($\mu$m) | Flux (mJy) | Error (mJy) | References |
|---|---|---|---|
| 8 | 0 | 150 | This work (IRAC) |
| 18 | 162 | 36 | This work (Akari) |
| 23.7 | 315 | 32 | Bryden et al. (2009) |
| 34.8 | 490 | 90 | Su et al. (2017) |

**Note.** The photometry at 8 $\mu$m falls 0.5% below the projected photospheric level (based on a model normalized to the 2–6 $\mu$m photometry), with an estimated error of 1%; we show it in Figure 1 as an error bar of 1% of the photospheric value from a level of zero.

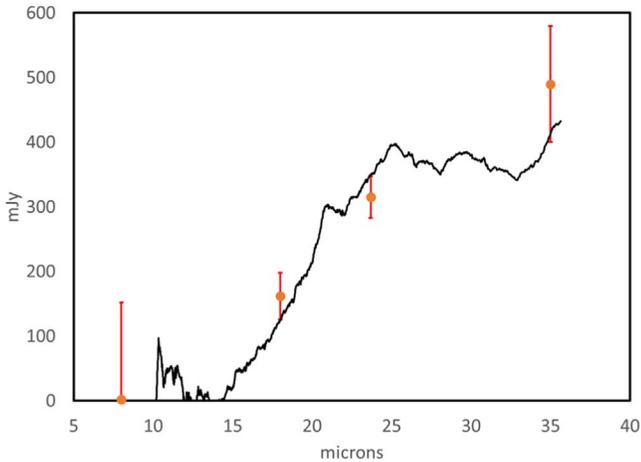

**Figure 1.** Photosphere-subtracted spectrum of the $\epsilon$ Eri inner and intermediate disks (Su et al. 2017), with photometry (red) provided in Table 2.

scaled with separate factors applied to the SH and LH modules to match the photosphere (SH; 1.09) and the MIPS 24 $\mu$m photometry (LH; 1.02) and finally smoothed with $R = \lambda/\delta\lambda \sim 30$. The SH module has a slit size of 4″7 [15 au] × 11″3 [36 au] with a 2″3 pixel$^{-1}$ plate scale, while LH has a slit size of 11″1 [36 au] × 22″3 [72 au] and a plate scale of 4″5 pixel$^{-1}$. Consequently, these data sets only contain flux contributed from the inner component of the debris disk material and are not representative of the debris rings resolved with ALMA at a diameter of ∼42″ (138 au; Booth et al. 2017).

The IRS spectrum (Su et al. 2017) implies that there is little excess emission above the stellar photospheric output at wavelengths shorter than about 15 $\mu$m. In fact, $\epsilon$ Eri is in a minority for weak output in the 10–18 $\mu$m range relative to other debris disks. For example, Jang-Condell et al. (2015) computed the signal-to-noise ratios (S/Ns) of the infrared excesses at three bandpasses (8.5–13, 21–26, and 30–34 $\mu$m) in the IRS spectra for a survey of 110 debris disk hosts, and found that the majority (62%) with significant detections ($\geqslant 4\sigma$) in the middle passband also show indications of excess ($\geqslant 1.5\sigma$) in the shortest one (8.5–13 $\mu$m). This effect is independent of spectral type; the 17 stars in this sample of type F5V and later also have an average excess indication $>1.5\sigma$.

It follows that the lack of excess emission at <15 $\mu$m seen in the $\epsilon$ Eri system may be unusual for debris disks of this age. To verify the spectrum and eliminate any possibility of systematic errors influencing its shape, we have assembled three photometric measurements of the debris disk (presented in Table 2). The longer two wavelengths are from the literature, but the derivation of the flux at 18 $\mu$m has been carried out for this work. All of these values apply to the inner 5″ radius, i.e., the inner and intermediate components—at 34.8 $\mu$m as determined by Su et al. (2017) and at the shorter two wavelengths under the assumption that the outer ring is too cold to contribute significantly.

To determine the disk output at 18 $\mu$m, we used Akari measurements that we self-calibrated for the ratio of fluxes at 9 and 18 $\mu$m for stars without infrared excesses but otherwise similar to $\epsilon$ Eri. Our initial sample was 28 stars measured at adequate S/N in both bands and with 4800 K $\leqslant T_{\text{eff}} \leqslant$ 6800 K. There was an indication of a trend with a larger ratio for the stars with $T_{\text{eff}} > 6000$ K. We dealt with this in two ways. In the first, we made a linear fit to the values of the ratio versus $T_{\text{eff}}$ and estimated the value at the temperature of $\epsilon$ Eri from this fit. In the second, we took just the stars with $T_{\text{eff}} < 6000$ K and averaged the ratios, since no trend was evident. The two approaches gave virtually identical estimates for the intrinsic ratio of $\epsilon$ Eri, 0.2174 ± 0.008, to be compared with the observed value of 0.2296 ± 0.0024. The ratio of the flux from the system to the flux expected with no excess is then 1.055 ± 0.012, from which we determined the disk emission.

As shown in Figure 1, the photometry agrees well with the IRS spectrum. This agreement removes any concerns about systematic errors in extracting the spectrum, such as in the slit loss correction, which is based on a point source. This result is important because the weakness of the emission in the 10–18 $\mu$m range is challenging to fit with standard models because of the strong silicate emission due to the 18 $\mu$m spectral feature in those minerals.

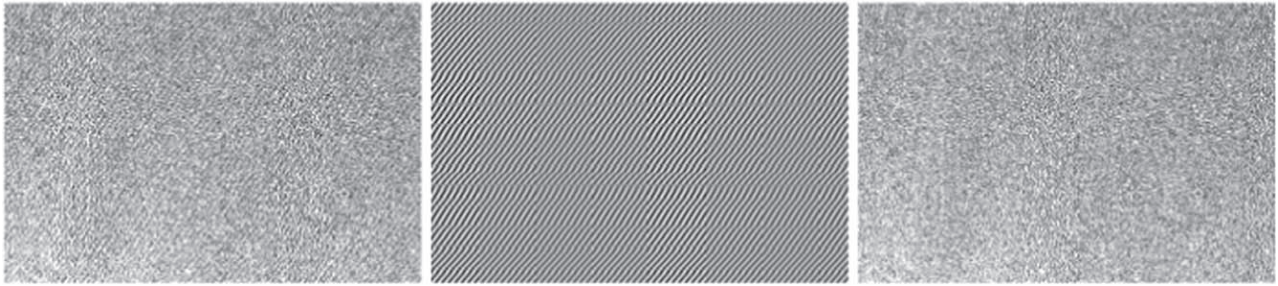

**Figure 2.** An example of the removal of the video-noise pattern from our STIS images, showed on a zoomed-in region of the detector. The amplitude of the noise pattern is around 1.5 ADU. Left panel: original RAW image; middle panel: noise pattern; and right panel: cleaned image.

### 3. Image Reductions

All of the HST observations were re-reduced using the calstis pipeline, available within the stistools package (part of Astroconda), maintained by the Space Telescope Science Institute. This was necessary to remove the video-noise pattern (also known as the "Herringbone pattern" for STIS) from the RAW data files, which is not part of the standard pipeline reduction.[4] Although the pattern is a minor contributor to the overall noise, to achieve the highest possible coronagraphic imaging contrast, we decided to perform these additional steps. The noise has an amplitude of around 1.5 analog-digital units (ADUs), which, as shown in Figure 2, would be visible in a 10.3 s exposure compared to the expected signal from faint and extended disk structures (which can be less than 1 ADU s$^{-1}$). This removal was performed using the Autofillet package (Jansen et al. 2003). In Figure 2, we show the effects of the video-noise on a RAW image frame, the pattern that is removed, and the final cleaned image. The calstis pipeline produces two calibrated images per exposure sequence, one with geometric distortions corrected (images ending with _sx2) and one without correction (images ending in _flt). As geometric distortions for STIS are minimal (less than 1 px at the edge of the detector; Walsh et al. 2001), and since our goal is to resolve extended emission and not a point source, we used the images without geometric distortion correction applied, as the corrections add an additional mathematical translation to the image processing, thereby—minimally—increasing its noise.

Following the standard image reduction steps, the coronagraphic observations were converted to units of counts per second (STIS pipeline processed images are in counts) before subtracting the PSF reference images from the science images.

The science and reference images were post-processed using three independent image reconstruction techniques. The first is reference differential imaging (RDI) wherein a set of reference PSF images is subtracted from a set of science images. The stable thermal environment and lack of atmospheric induced seeing provide a relatively stable PSF. When combined with a set of PSF reference images observed concurrently with the science images, STIS+RDI is capable of producing contrasts[5] deeper than $\sim 10^{-10}$ at wide separations (e.g., Schneider et al. 2016) using classical PSF subtraction techniques alone. However, additional challenges for this PSF subtraction arise because of the nearly face-on orientation of the $\epsilon$ Eri disk.

Angular differential imaging techniques that leverage azimuthal diversity in observed disk morphology are not helpful, and the radial disk profile can be misrepresented as a change in the slope of the PSF.

Ground-based adaptive optics systems have produced galleries of debris disk images in scattered light (e.g., Esposito et al. 2020) utilizing more complex image reconstruction algorithms, though these data sets have very different temporally varying noise properties than our HST data. In general, classical reference PSF subtraction is better than principal component analysis (PCA) approaches like Karhunen–Loéve Image projection (KLIP; Soummer et al. 2012) for STIS data sets, especially for faint disks, largely because it prevents oversubtraction. Nonnegative matrix factorization (NMF; Ren et al. 2018, 2020) is a similar approach but exploits the fact that a disk signal should be positive. Theoretically, this better handles the slope caused by differences in science and reference images. NMF has been shown to improve the signal-to-noise of a disk detection, though classical PSF subtraction still performs best for the faintest disks observed with STIS coronagraphy. (See further discussion in Ren et al. 2018.) Below, we detail our results using these three techniques: classical PSF subtraction (classical RDI), KLIP, and NMF.

#### 3.1. Method 1: Classical RDI

Classical RDI is the simplest coronagraphic post-processing method and has been proven to be reliable and effective for systems where the PSF of the telescope optical tube assembly (OTA) is stable, such as for a space telescope, like HST. In simplest terms, the PSF of the reference source is subtracted from the observed target images, and the "clean" images are combined, using masks where necessary to ignore contributions from imaging artifacts such as the OTA diffraction spikes and coronagraphic apertures. Improvements within classical RDI processing can be made, by combining sequences of images prior to translations and/or PSF subtractions, and by selecting the optimal sequence between image translations and PSF subtractions, the smoothest translation functions, and the optimal scaling of the reference PSF (and even the target source in some cases). We tested a few combinations of these variables for the classical RDI processing of our GO15906 data set. Details of the process and a description of the approach that yielded the highest contrast—relative to $\epsilon$ Eri—final image can be found in Appendix B.1. This includes a description of the image centering, changes in science/PSF scaling due to OTA "breathing," and final image combination.

We show the final images of our classical RDI post-processing in Figure 3. All of the images employ $3\sigma$ clipping around the data median for both combination steps (per epoch and final all epochs); however, they use either median or mean averaging for the individual combination steps. The image with the least number of artifacts is the "median of medians," where both the per-epoch observations and also the final translated images were combined with median averaging. Visually, this image is the most pleasing, although the "mean of means" image appears to be the smoothest outside of apparent artifacts.

---

[4] The pattern is present in all images taken after 2001 July, when the instrument was switched over to its Side 2 electronics, due to the failure of its primary electronics on Side 1.
[5] Contrasts discussed in this work and in Schneider et al. (2016) are computed as a $1\sigma$ median absolute deviation (MAD) for a PSF-subtracted data set.

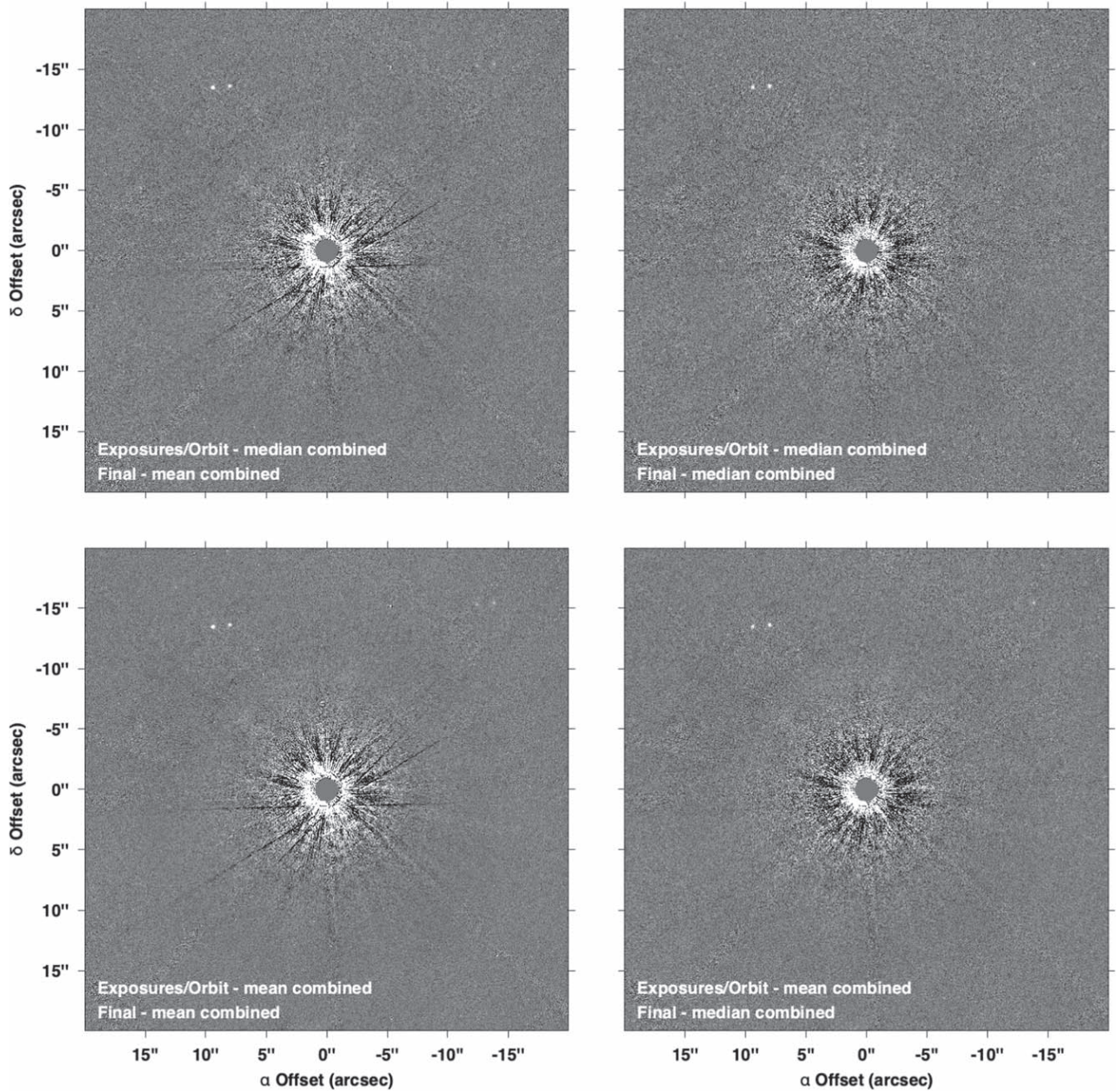

**Figure 3.** The results of the classical RDI technique, shown here using various averaging methods, all employ $3\sigma$ clipping around the data median. The best contrast curve is achieved in the "mean of means" combination (lower left image); however, it shows some leftover imaging artifacts from masking. The median of medians is the smoothest, and its contrast curve (Figure 4) is not considerably worse than the best case. The images are shown using a linear scale between $-1$ and $+1$ counts s$^{-1}$.

We determine the imaging sensitivities for all of our final images, as in Figure 4. As the goal of our program was to resolve the debris disk around $\epsilon$ Eri spatially, we use the median absolute deviation (MAD) as a metric of residuals instead of the commonly used standard deviation. The MAD is defined as the median of the deviations around the data median, i.e.,

$$\mathrm{MAD} = \mathrm{median}(|x_i - \tilde{x}|), \quad (1)$$

where $\tilde{x}$ is the median of all $x_i$. The MAD is relatively robust to outliers, especially for smaller data sets; it is therefore the better estimator for our case (see, e.g., Schneider et al. 2014). For larger samples, $1\sigma_{\mathrm{SD}} \approx 1.48$ MAD. We determined the value of the MAD in 2 px wide annuli, centered on $\epsilon$ Eri. The lowest (best) sensitivity threshold is achieved using $3\sigma$ clipping around the data median, although $2\sigma$ clipping yields similar results.

We show the radial surface brightness profile (median averaged over 360°) in Figure 5. As the figure shows, inside a radius of $\sim 2''$, systematic errors associated with the strong residual signal from the star are dominant and are not necessarily fully reflected in the MAD. The left panel in the figure shows that we were unable to detect any extended surface brightness features outward of $5''$ in our data set. In the inner regions, however, we see a very tentative detection between $1''\!.2$ and $1''\!.7$, which corresponds to 3.84 and 5.44 au, just outside the orbit of $\epsilon$ Eri b.

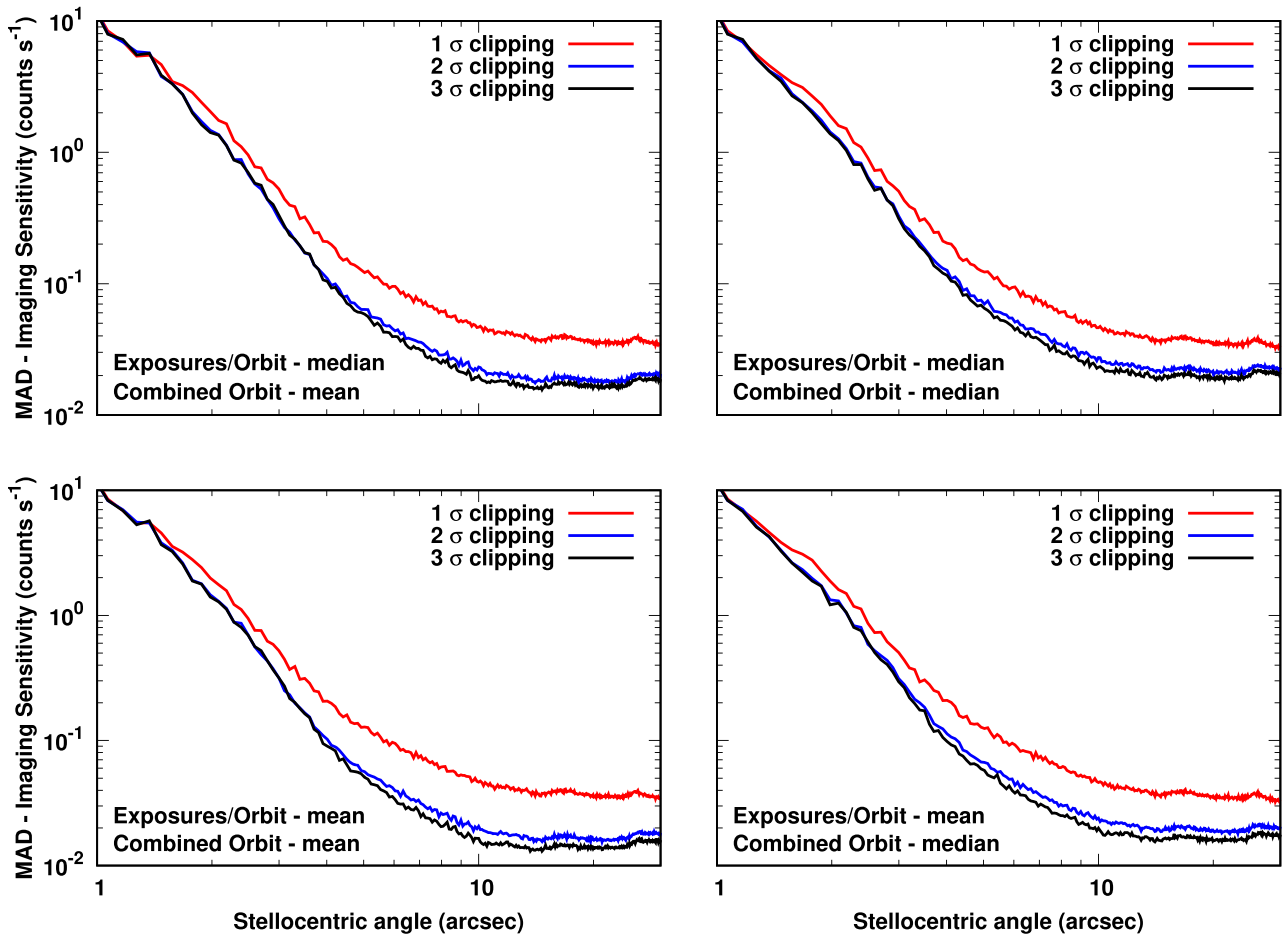

**Figure 4.** The median absolute deviation (MAD) detection sensitivity curves achieved by various averaging methods for the entire eight orbit data set, shown here in counts per second. The averaging methods were varied for the combination of all data within an orbit and also for the final combination of images produced per orbit. The figure shows that $3\sigma$ clipping of data around the median produced the best imaging sensitivity in all cases. While the mean combination of images yields a slightly better overall sensitivity than the median combinations, as shown in Figure 3, the median combinations successfully remove certain artifacts and yield a slightly more even background level.

In Figure 6, we show the MAD contrast relative to the brightness of $\epsilon$ Eri achieved with our reductions. We compare this contrast to a Mie-scattering model based on the best-fit dust distribution in Su et al. (2017) and find we should have achieved a strong detection. The model included belts from 1.5–2, 8–20 and 52–76 au with a grain size distribution exponent of 3.65, a minimum grain size of 0.5 $\mu$m and opacities from Ballering et al. (2016; see Section 4 for a more detailed description). We also compare the achieved contrast to the levels of other STIS programs. According to the STIS ETC, an aperture containing 100% of the flux from $\epsilon$ Eri would yield $9.27 \times 10^8$ counts s$^{-1}$ measured flux at our GAIN = 4 setting. We divided our observed MAD contrast curve by this value to calculate our observational limits for extended sources. We compare this contrast curve to the one derived from the MAD background sensitivity curve shown in Figure 4 of Schneider et al. (2016) for HD 207129, which is also a later-type star. We converted the Vmag arcsec$^{-2}$ units to instrumental units, using pysynphot, assuming a 5000 K stellar spectrum, which estimates 1 count s$^{-1}$ to equal 24.4931 Vega mag. We then divided the instrumental MAD sensitivity curve with the instrumental brightness of HD 207129, obtained the same way with the STIS ETC as we did for $\epsilon$ Eri. The contrast curves for the two later-type stars match remarkably well. To see how our programs compare to likely the deepest combined STIS data set so far, we also plot the MAD contrast curve for the Fomalhaut image shown in Gaspar & Rieke (2020), which combines almost 8 hr of total integration time on the star from multiple individual programs. As Fomalhaut is an early-type star, its PSF will be considerably narrower, resulting in a steeper contrast curve. The $\epsilon$ Eri and HD 207129 observations are contrast limited to around $8''$, outward of which the photon noise dominates. The deep Fomalhaut data set is contrast limited to $15''$–$20''$.

### 3.2. Method 2: Karhunen–Loéve Image Projection

In addition to the classical PSF subtraction method performed above, we also employed a KLIP Algorithm (Soummer et al. 2012). The same PSF reference files are used, but instead of subtracting the full stack (coadded images obtained at the same aperture per each visit), the individual exposures are used to create an orthogonal basis set of eigenimages using a KL transform onto which the science images are projected. This is one in a family of techniques that constructs a synthetic reference PSF from a linear combination of reference images, but in this case the number of KL modes is truncated and only the most correlated modes are used in the final subtraction. Increasing the number of modes used will allow for probing deeper contrasts at closer inner working angles, but will also remove fainter or more extended features. KLIP has been shown to improve the achieved contrast for HST coronographic observations in the case of high-S/N disks (e.g., Debes et al. 2019; Ren et al. 2019a). However, it is also prone to oversubtraction, especially in disks viewed near-face-on where the extended disk signal is mistaken for residual PSF structure and removed.

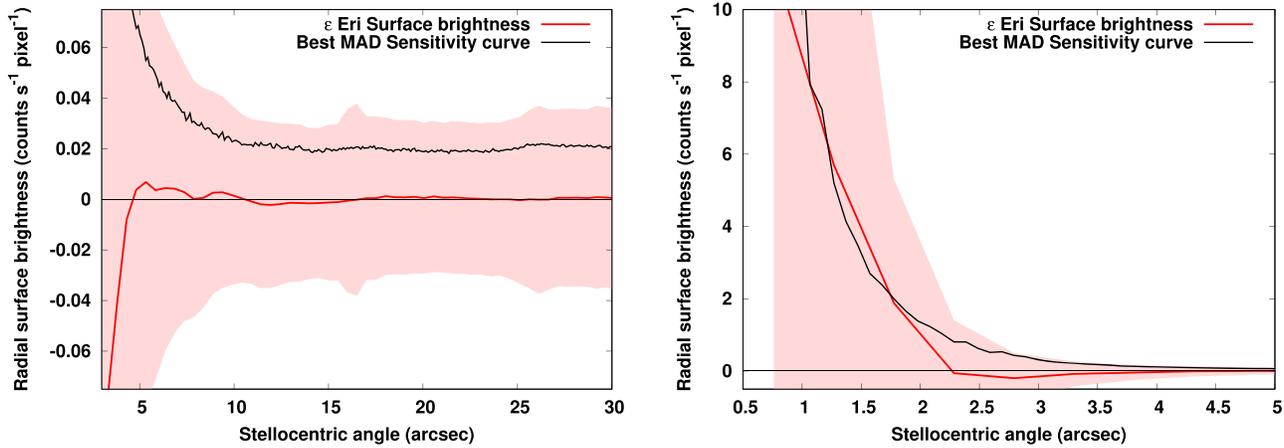

**Figure 5.** The radial surface brightness profile of our classical RDI reduction, shown here with 1$\sigma$ standard deviation per pixel error bars in red and the MAD sensitivity curve shown in black. Extended sources can be detected below the 1$\sigma$ level for point sources because we average the MAD over a two pixel width annulus. The detection level is even lower for extended components that span a number of these annuli. Left panel: the radial surface brightness curve between 5″ and 30″. While positive bumps are present at around 5″ and 10″, they are well below the detection threshold. The MAD detection threshold at 10″ is around $\approx$0.020 counts s$^{-1}$ and the profile at this radius is $\approx$0.003 counts s$^{-1}$. Right panel: the radial surface brightness curve between 0″.5 and 5″, with a very tentative detection between 1″.2 and 1″.7. The uncertainty in this result is dominated by systematic errors arising from the PSF subtraction.

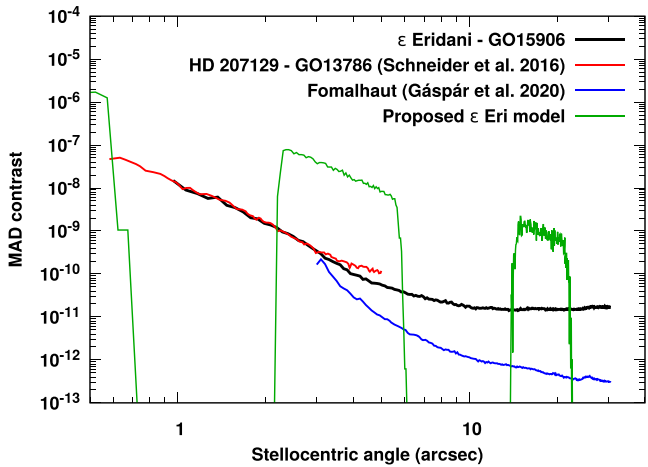

**Figure 6.** Comparison of our initial (proposed) model (based on that of Su et al. 2017, but with $a_{\min} = 0.5$ $\mu$m) and the MAD contrast curve achieved with classical RDI within our GO15906 program and to those in GO13786 (Schneider et al. 2016) of HD 207129 (also a late-type star) and Fomalhaut (Gaspar & Rieke 2020), an early-type star. The contrast curves of the two late-type stars agree remarkably well. Our $\epsilon$ Eridani observations are contrast limited to around 8″, from which point on, photon noise dominates outwards. The Fomalhaut observations are significantly deeper (27,402 s in total, compared to our 3141 s) and are of an earlier spectral-type star, which will have a narrower PSF. In addition, the Fomalhaut observations used roll-differential imaging (Fomalhaut was its own PSF), which provides the "perfect" color-matched PSF. Taking these into account, it is remarkable that we achieve similar contrasts at 3″, and not surprising that the Fomalhaut observations do better farther out.

We performed the calculation using the pyKLIP package (Wang et al. 2015) using a truncation of 5 KL modes to optimize for extended structures.[6] Further details of the reduction can be found in Appendix B. The combination of individual exposures and the image centering were performed differently for KLIP than for the classical RDI case above. These changes along with a more detailed algorithm description are shown in Appendix B.2.

Figure 7 displays the resultant images, while Figure 8 displays the achieved MAD contrast values and the image sensitivities. The results are very similar to the classical PSF subtraction case, but with slightly worse contrasts near the inner working angle and a steeper slope out to $\sim$10″ followed by a plateau. This technique was also not able to retrieve the intermediate disk around $\epsilon$ Eri.

### 3.3. Method 3: Nonnegative Matrix Factorization

In a final attempt to retrieve the $\epsilon$ Eri disk signal, we employ vectorized NMF. NMF is similar to KLIP in that it uses a basis set of components generated from reference images to model the target image. Unlike KLIP, the basis set created in NMF is both nonorthogonal and nonnegative. Since any disk or planet signal is inherently nonnegative, this method is less prone to oversubtraction than KLIP/PCA analysis. The NMF algorithm and further details of this post-processing analysis are presented in Appendix B.3.

Figure 9 presents the NMF result. Note that the image scaling is presented logarithmically here to capture the larger dynamic range of the NMF result as compared to the classical RDI or KLIP results. The corresponding MAD contrast is presented in Figure 8. While NMF is able to match the contrast performance of classical RDI outside of $\sim$12″, it underperforms the other two methods at closer separations. Attempts to improve this result by changing the number of components or employing best factor finding (see Ren et al. 2018) were unsuccessful. Without any strong disk signal, it is likely that NMF attempts to represent any changes in the slope between the science and reference images as positive disk signal. While it has been proven that NMF can provide a stronger S/N detection of bright disk images with the STIS coronagraph (Ren et al. 2018), the $\epsilon$ Eri debris disk is too faint for a favorable NMF result.

---
[6] We tested several values for the KL mode truncation parameter, and 5 was chosen to be a compromise between achieving a clean PSF-subtracted image while minimizing oversubtraction.

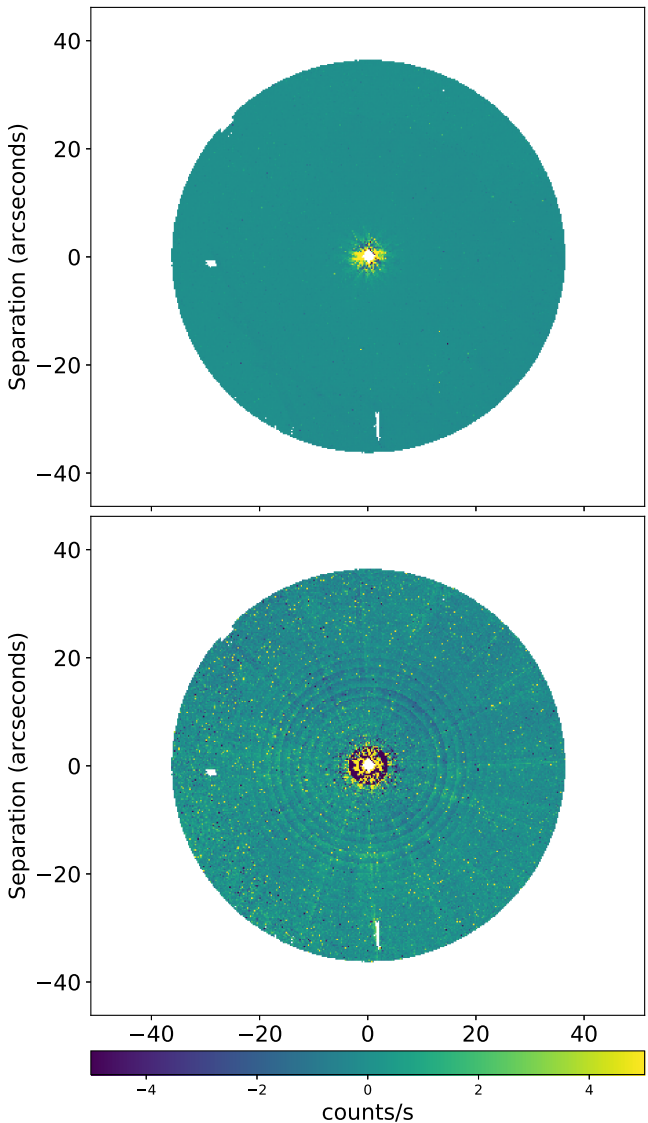

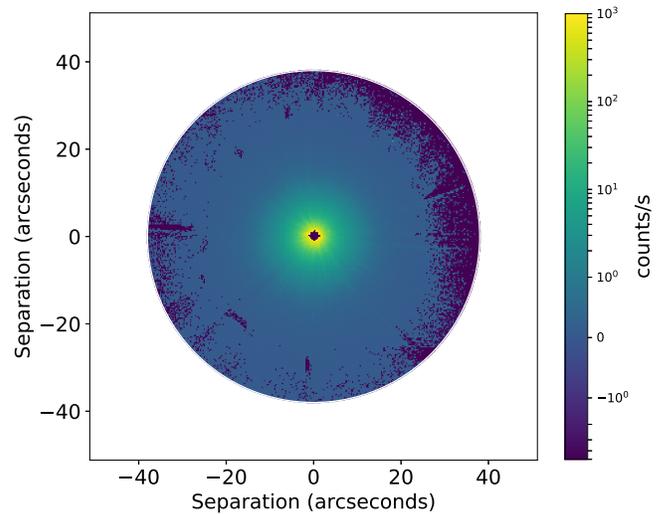

**Figure 9.** The resultant images for the NMF subtraction technique on the $\epsilon$ Eri data set performed with a single zone using five components to maximize the diffuse disk signal. The image is shown with north pointing up with an IWA of $1''\!.4$ and an OWA of $35''$.

### 3.4. Summary of Coronagraphic Results

The intermediate disk for $\epsilon$ Eri was not detected in our data set; however, we are able to place a strong constraint on the scattered light surface brightness from any small dust grains in the 3–40 au ($1''$–$12''$) region. The outer, Kuiper Belt analog imaged with ALMA was also not detected. As shown in Figure 8, the achieved contrast provides a surface brightness sensitivity floor around 4 $\mu$Jy arcsec$^{-2}$ at radial locations outside of $10''$, where the photon noise limit is reached. We choose to use the surface brightness limits provided by the classical PSF subtraction method to test against the $\epsilon$ Eri disk models. Given the unique stability and high sensitivity to low surface brightnesses of HST, classical PSF subtraction often provides the best results for extended structures. For brighter detections, PCA methods like KLIP can increase the S/N, but for faint extended structures, the potential for oversubtraction can wash out the disk signal and make the results more difficult to interpret. In any case, neither of these methods produced a better result than classical PSF subtraction.

### 4. Data Analysis and Modeling

Here we combine this scattered light nondetection with data provided in the literature to fine tune a disk model for the $\epsilon$ Eri system. Numerous models have been introduced to explain the SED of the system (see below). Generally, these models include a warm inner component with a sizable small dust grain population (likely interacting in some way with the planet b) and a cold outer belt analogous to our own Kuiper Belt comprising mainly larger planetesimals. The outer belt has been resolved by a number of groups (e.g., Greaves et al. 2014; MacGregor et al. 2015; Chavez-Dagostino et al. 2016). Recent ALMA observations show it to be a narrow ring centered at 69 au with a width of 11–13 au and an inclination of 34° (Booth et al. 2017). These authors do not provide a mass estimate for the outer belt, though their results are in agreement with previous observations of the outer belt at lower resolution (Greaves et al. 2014; Chavez-Dagostino et al. 2016).

**Figure 7.** The resultant images for the KLIP PSF subtraction technique of the $\epsilon$ Eri data set performed with a single zone (top) and 20 subannuli (bottom). In both cases, we used 5 KL modes to maximize the diffuse disk signal. Both images are shown with north pointing up with an inner working angle (IWA) of $1''\!.4$ and an outer working angle (OWA) of $35''$.

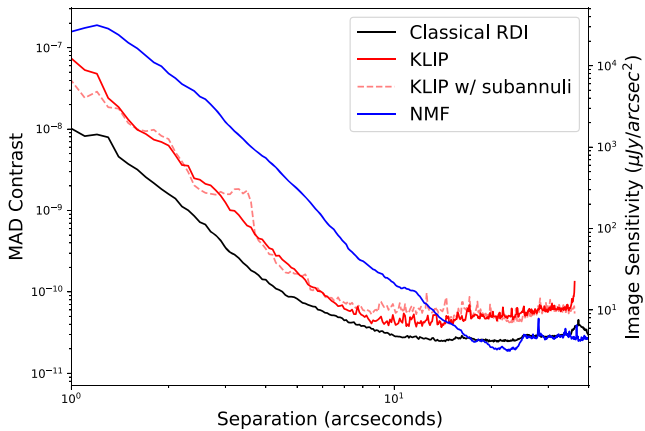

**Figure 8.** The MAD contrast curves for the KLIP and NMF results as compared to the classical PSF subtraction. Image sensitivities are shown on the right. Classical RDI subtraction provides the deepest contrasts over the full range in separations.

In addition, it is believed that a warm, intermediate dust system is located in either (1) one or more planetesimal belts; or (2) small grains dragged in from the cold outer belt due to PR drag. For a more complete discussion of these two modeling families, see Su et al. (2017), who concluded that the first hypothesis is more likely. This conclusion is reinforced by the lack of an excess in IRAC Band 4 (6.4–9.3 $\mu$m; see Figure 1), since the second class of model predicts a significant flux from these dragged in grains when they approach close to the star (Su et al. 2017).

Various locations have been suggested for the intermediate belt, but in general, they all predict a significant amount of dust in the zone where our coronagraphic observations are most sensitive. Spitzer observations initially suggested a dust belt at 20 au with a broad outer disk at ~35–90 au and a tentative very warm component at 3 au (Backman et al. 2009). This picture was revised based on the addition of SOFIA data to an intermediate belt from 8–20 au in Su et al. (2017). Herschel/PACS results are interpreted in terms of dust belts with radial extents of 13–21 and 36–72 au (Greaves et al. 2014). Chavez-Dagostino et al. (2016) required a component inside the outer belt, either as a ring at 30 au or a broad belt with a flat surface density from 14–63 au. Assuming that these belts are generating small dust by collisional cascades, e.g., down to sizes of $\lesssim$0.1 $\mu$m (e.g., Krivov 2007), nearly all would produce a significant level of fine dust within ~30 au that should scatter in the HST band significantly. In the case where the warm dust is produced by small grains pulled in from the outer disk by PR drag (Reidemeister et al. 2011), the smallest grains will spiral until they reach ~4.3 au where they will stall because of the effect of the planet. Again, these models produce a significant population of very small (~0.05 $\mu$m; Reidemeister et al. 2011) scattering grains within 30 au.

To confirm that the leading literature models for the $\epsilon$ Eri debris disk system would have been detected in our STIS observations, we first ran a simple test. We simulated the most probable model established in Su et al. (2017) using the code DUSTMAP (Stark 2011). We placed the intermediate belt between 8 and 20 au. The outer cold component lies between 52 and 76 au based on previous radio data (Greaves et al. 2014; Chavez-Dagostino et al. 2016). For the particle size distribution (represented by $n(a) \propto a^{-p}$), we assumed a slope coefficient of $p = 3.65$, a minimum size of $a_{\min} = 0.5$ $\mu$m, and a silicate- and organic-rich grain composition as determined by Ballering et al. (2016). We find that, if the properties of the intermediate belt from 8–20 au were as predicted, the disk would have been easily detected with HST based on our achieved contrast as demonstrated in Figure 6. This is the same dilemma encountered by many previous studies combining models of scattered and thermally emitted energy from debris disks (e.g., Krist et al. 2010; Golimowski et al. 2011; Lebreton et al. 2012).

To overcome this tension between the thermal disk models for $\epsilon$ Eri and our scattered light nondetection, we must update the model to reproduce the dust content implied by the infrared measurements while decreasing the scattered light surface brightness. We consider the possibilities below:

1. *The Inner Disk and Epsilon Eri b.*
    We first investigate the dust content that could be hidden inside of the inner working angle (<1″) of our STIS data set. The SED provides some constraints on the dust population very close to the central star. From Section 2.2.1, there is very little infrared excess at wavelengths short of 8 $\mu$m. Recent results from the HOSTS survey probe for exozodiacal dust at 11 $\mu$m in the habitable zone (roughly equivalent to the inner ~0″.6 region of $\epsilon$ Eri) and find $297 \pm 56$ zodis (where 1 zodi is the vertical geometrical optical depth of dust in the habitable zone of the solar system; Ertel et al. 2020). Furthermore, the authors find that the excess is larger with increasing aperture size, implying that the dust may extend farther outward. However, the IRS spectrum (Su et al. 2017) shows that there is only a low level of emission from circumstellar material out to ~15 $\mu$m. Further complicating the picture is the presence of a planetary companion, $\epsilon$ Eri b, with a semimajor orbital axis of 3.48 au, a mass of $M_b = 0.66^{+0.12}_{-0.09}$ $M_{\rm Jup}$, and an inclination of $i = 77°.95^{+28.50}_{-21.06}$ (Llop-Sayson et al. 2021). Given the best-fit orbital parameters, Llop-Sayson et al. (2021) predicted that the disk will be cleared from 2.97–4.29 au. Consequently, while there may be an exozodiacal dust population interior to $\epsilon$ Eri b, the STIS observations are not sensitive to this material with an IWA of ~1″ (3.5 au). However, hiding enough dust in this region to contribute to the infrared SED is challenging given the constraints imposed by the IRS spectrum discussed below and the lack of excess at wavelengths at 8 $\mu$m and shorter as discussed in Section 2.2.1.

2. *The Intermediate Disk and Constraints from IRS Spectroscopy.*
    The IRS spectrum places the strongest constraints on the warm dust debris disk components inside of $4″.7 \times 11″.3$ for the 10–20 $\mu$m data and $11″.1 \times 22″.3$ for the 19–37 $\mu$m data (see Section 2.2). These data show a steep rise in the IR excess from 15–25 $\mu$m followed by a plateau from 25–35 $\mu$m. Previous fits to the IRS spectra tend to lie above it in the 13–20 $\mu$m range, but our newly derived point at 18 $\mu$m verifies that the placement of the spectrum is correct. We examine if an ensemble of grains is capable of reproducing the shape of the IRS spectrum in Section 4.2.3.

3. *The Outer Disk.*
    Despite the extreme image sensitivities achieved at this separation (~4 $\mu$Jy arcsec$^{-2}$), the STIS coronagraphic data set does not detect a scattered light ring colocated with the planetesimal belt observed with ALMA at 69 au (21″). Small grains produced by collisions in the planetesimal belt could have a relatively low surface brightness, since PR drag will distribute them over a large range of radii. Additionally, a combination of roll angle availability and the STIS coronagraphic field of view contribute to a sparse science pixel coverage at these separations. Proffitt et al. (2004) failed to detect the outer belt in nominally deeper visible-light observations, although the effect on this result of roll angle cancellation is difficult to evaluate. The broadband SED can be used

to place additional constraints on the location of the smaller dust grains in the system (see Section 4.2.4).

### 4.1. Dynamical Modeling of the ϵ Eridani System

In these next sections, we combine constraints from the scattered light observations, the IRS spectrum, the broadband SED and the system dynamics to improve our understanding of the debris and planetary system surrounding ϵ Eri. To that aim, we construct a dynamical model of the system that includes both ϵ Eri b (3.5 au; Llop-Sayson et al. 2021) and the putative planet c with a mass of 0.3 $M_{Jup}$, which is hypothesized to be responsible for the narrowness of the outer planetesimal belt (44 au; Booth et al. 2017).

Dynamical and radiative transfer modeling were performed using the code DiskDyn (Gaspar & Rieke 2020). DiskDyn evolves the spatial distribution of dust and planetesimals in a planetary system, including the effects of gravity—from both the host star and planets—and stellar radiative forces. Additionally, it also calculates the thermal and scattered light emission from the dust particles as the system dynamically evolves for dusts of various optical properties. We initiated our dynamical modeling with a continuous disk from 1.5–100 au, a disk opening angle of 5° (vertical Gaussian distribution), a standard deviation of 0.1 in eccentricities, and a surface density profile of $r^{-1}$. The largest bodies in the disk were 100 m in radius, while the smallest sizes were varied for the fitting. The disk contained a total mass of 0.3 $M_\oplus$, distributed among 5 million tracing particles. The emission profiles were calculated for particle radii of 0.1 $\mu$m to 1 cm, in 363 logarithmically distributed sizes. The dynamical model was evolved to 160,000 yr.

We assume a 25° inclination for the debris disk. It is expected that the integrated photometry is agnostic to the viewing angle of the disk. Conversely, the more inclined the disk, the easier it is to detect in scattered light as both (i) the line-of-sight optical depth increases and (ii) the geometry approaches the forward scattering peak of the dust grains. This near-face-on orientation is conservative from a surface brightness standpoint and is coplanar with the outer cold belt (Booth et al. 2017) and with the stellar axis of rotation (Campbell & Garrison 1985; Saar & Osten 1997). However, it differs from the current best estimate of the inclination for the orbit of ϵ Eri b.

The resultant radial particle distributions for the dynamical clearing of a homogeneous disk via the two planets in the system are presented in Figure 10. As expected, ϵ Eri b is efficient at clearing dust colocated with and interior to its orbit, an effect that is less pronounced for larger particle sizes. Likewise, ϵ Eri c clears dust in its orbit but also affects some material in resonant orbits. ϵ Eri c is also responsible for shaping the Kuiper Belt analog with the inner and outer disk edges coinciding with the 3:2 and 2:1 mean motion resonances, respectively. The dust density falls off steeply outside of ∼100 au. There is a slight buildup of the smallest dust particles outside of ϵ Eri b, which corresponds to the location of the very tentative detection illustrated in Figure 5. This feature is present (though at a lower magnitude) for the 1–10 $\mu$m dust population and, if real, will be visible in planned James Webb Space Telescope (JWST) observations. The dust distribution is largely unaffected by the dynamical effects of the two assumed planets in the region from 7–40 au. This does not imply dust must be present at these separations in the ϵ Eri system, but rather that any dust depletion in this region must be caused by additional planets or by some other mechanism in the primordial disk phase.

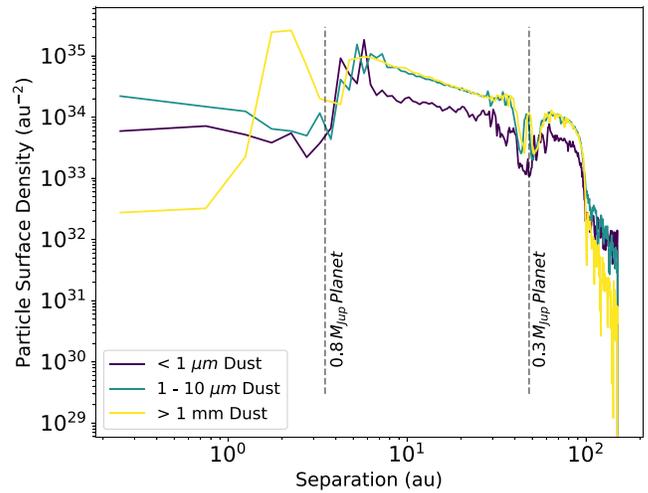

**Figure 10.** The normalized particle number density as a function of separation is presented for a dynamical model of the ϵ Eri circumstellar system. The influences of the two planets on the dust distribution is strongly dependent on the grain sizes. The 1–10 $\mu$m and >1 mm dust have been scaled up for illustration using values of $3 \times 10^2$ and $3 \times 10^{10}$, respectively. The dust distribution from ∼8–35 au resembles the input radial density, since the model does not include planets at radii that would interact strongly with it. That is, in this range there are no strong constraints on the ring structure imposed by the dynamical model.

### 4.2. Photometric Constraints via Radiative Transfer Modeling

Debris disk models based only on SEDs derived from photometry are notoriously degenerate, most notably, between grain sizes and radial location. However, in this case we have (1) a very detailed set of SED constraints, including the IRS spectrum bridging from 10 to nearly 40 $\mu$m, (2) sufficient spatial resolution to separate the inner and outer zones of the debris system, (3) the lack of detection in scattered light, placing limits on the population of small dust grains, and (4) the dynamical modeling to identify zones that should have significantly reduced dust densities. The purpose of this section is to explore the degeneracies and conclusions that can be drawn from SED models.

We use a radiative transfer code with a Markov Chain Monte Carlo (MCMC) framework to fit against the observables. The chosen dust properties are presented in Section 4.2.1. Next, we outline the modeling approach itself in Section 4.2.2. Given the difficulty in separating the radial dust contributions to the IRS spectrum and the broadband SED, we chose to fit against these two observables separately as proxies for the intermediate and outer disks, respectively, with results presented in Sections 4.2.3 and 4.2.4. Appendix C provides an additional exploration into what insights to the disk structure would be possible if we had only the broadband infrared photometry that is typical for the majority of debris disks at much greater distances than ϵ Eri.

#### 4.2.1. Assumed Dust Properties

We use dust opacities presented in Ballering et al. (2016) designed to comprehensively fit resolved images of the β Pictoris debris disk from optical to radio wavelengths. Their model includes 60% astronomical silicates and 40% organic refractory material, while disfavoring porous or icy grains, and reduces the scattered light signature by a factor of 5 compared to pure astronomical silicates. This is similar to the composition derived for the HR 4796A disk (Rodigas et al. 2015) that preferred silicates and organics over water ice. We note that both β Pictoris and HR 4796A are more luminous than ϵ Eri; however, the debris disks probed in those studies were located farther from their host stars with comparable dust temperatures to the intermediate disk regions of ϵ Eri. Furthermore, the grain composition recommended by Ballering et al. (2016) does not include volatiles, so the temperature regime may not be critical. Although more extreme compositions with lower albedos could further reduce the scattered light, pushing farther in this direction to explain the low level of scattered light in the ϵ Eri system will drive toward less plausible grain compositions on the basis of standard abundances.

The minimum dust grain size in the system is set by the radiation pressure blowout size. Recent work by Arnold et al. (2019) shows that the blowout size is dependent both on the grain composition and shapes; it derives estimates for the minimum grain size expected for the ϵ Eri disk. For compact spheres using the Lorenz–Mie theory, the blowout size ($s$) is unconstrained for astronomical silicates and ices, i.e., it is predicted that all grain sizes will be stable against blowout. However, $s = 0.4$ $\mu$m for pure amorphous carbon, and $s = 0.3$ $\mu$m for a mixture of 33% amorphous carbon with silicates and ices. For more realistic particle morphologies (76.4% porous agglomerates modeled using the discrete dipole approximation), the blowout sizes remain unconstrained for astronomical silicates and ices but increase to ∼1 $\mu$m for amorphous carbon and carbon-heavy mixtures. Likewise, extremely porous grains (97.5%) achieve even larger blowout sizes of up to 4 $\mu$m. That is, realistic grain properties may result in evacuation of the intermediate region of submicrometer grains via photon pressure force blowout. We included models with minimum grain sizes above $s$ both to allow for such properties and to include the possibility that other effects may limit the production of very small grains (Pawellek & Krivov 2015). The minimum grain size is left as a free parameter in our modeling.

Theoretical modeling suggests that collisionally dominated debris disks in a quasi–steady state can have a range of particle size distributions, $n(a) \propto a^{-p}$ with the index $p$ ranging from, e.g., 3.5 for the classical case (Dohnanyi 1969) to the steeper 3.65 (Gáspár et al. 2012). The slope of the ϵ Eri long-wavelength (>100 $\mu$m) SED indicates a relatively small value of $p$ (e.g., Gáspár et al. 2012). In practice, the SED slope provides a poor constraint on the underlying size distribution due to uncertainties in the dust optical properties at long wavelengths (Löhne 2020) and outside dynamical effects can remove small dust grains from the system (e.g., Pawellek & Krivov 2015) allowing for shallower size distribution slopes. The grain size distribution slope was left as a free parameter in our models with values ranging from $p = 3.4$–3.8.

### 4.2.2. Approach to Modeling

For a quick reconnaissance of possible belt locations and dust properties, we used Debris Disk Simulator (Wolf & Hillenbrand 2005). We found that it is difficult to achieve a fit with a single particle size distribution. First, a relatively steep particle size distribution is critical to suppress the emission at the longer wavelengths and thus to fit the flat behavior of the spectrum for the intermediate region from 20–35 $\mu$m. A potential issue is that a large population of dust grains in the intermediate region extending to submicrometer sizes would likely have been detectable in the STIS data set. Consistent with the results from Löhne (2020), a more shallow particle size distribution is needed for the outer ring detected by ALMA.

This manual approach does not allow for a comprehensive exploration of dust particle properties and in general cannot determine a full probability distribution for the allowed fits. To probe this degenerate parameter space more thoroughly, we used a more complex model to explore the tension between scattered and thermally emitted flux in the intermediate zone that is illustrated in Figure 6. Building on the results from the dynamical model in Section 4.1, we split the contributions from the resultant dust populations into radial bins and used a radiative transfer code assuming Mie-scattering theory and dust optical constants from Ballering et al. (2016) to generate a library of SEDs. We chose 13 zones with an uneven sampling to capture the features of the dynamical model shown in Figure 10. The innermost radial bin lies within the orbit of ϵ Eri b, while the largest has a width of 30 au in the outermost regions of the disk past the unseen (in scattered light) ring at 69 au. The full list of radial bin sizes is shown in Table 3. The stellar photospheric SED was scaled to the model SEDs for wavelengths below ∼2 $\mu$m, where disk emission is negligible, and subtracted. The model SEDs were then resampled to the resolution of the observed data set.

Table 3
MCMC Disk Results

| Radial Zone | IRS Best Fit | | SED Best Fit | |
|---|---|---|---|---|
| (au) | Mass ($M_{Earth}$) | $\sigma$ (m$^2$) | Mass ($M_{Earth}$) | $\sigma$ (m$^2$) |
| 0–3 | 9.81e-08 | 2.82e+18 | 1.21e-06 | 6.35e+18 |
| 3–4 | 2.37e-09 | 1.76e+17 | 1.17e-09 | 8.56e+15 |
| 4–6 | 2.40e-08 | 4.74e+18 | 1.41e-07 | 1.10e+18 |
| 6–12 | 7.62e-08 | 1.09e+19 | 1.08e-07 | 8.85e+17 |
| 12–20 | 2.65e-08 | 3.80e+18 | 1.16e-06 | 9.61e+18 |
| 20–30 | 2.97e-05 | 4.41e+21 | 1.10e-05 | 9.13e+19 |
| 30–37 | 2.17e-07 | 3.10e+19 | 5.36e-04 | 4.39e+21 |
| 37–44 | 2.16e-08 | 2.35e+18 | 1.14e-06 | 9.03e+18 |
| 44–52 | 7.10e-07 | 9.11e+19 | 4.03e-03 | 3.31e+22 |
| 52–63 | 1.29e-07 | 2.06e+19 | 1.13e-04 | 9.34e+20 |
| 63–85 | 5.92e-07 | 8.96e+19 | 8.35e-06 | 6.98e+19 |
| 85–100 | 1.91e-07 | 3.01e+19 | 5.02e-05 | 4.22e+20 |
| 100–130 | 1.52e-07 | 9.12e+19 | 2.99e-07 | 3.23e+18 |
| Sum of Zones | 3.2e-05 | 4.8e+21 | 4.8e-03 | 3.9e+22 |
| Dust Particle Properties | IRS best fit | | SED best fit | |
| Size Dist. ($p$) | 3.8 | | 3.4 | |
| $a_{min}$ ($\mu$m) | 0.5 | | 2.0 | |

**Note.** We present the parameter values for the best-fit model MCMC results optimized for the IRS spectrum (left) and the outer disk SED (right). Mass and particle scattering cross sections for each radial zone (as described in Section 4.1) are computed from the best-fit scale factors. Masses are computed using the best-fit dust properties listed and a maximum particle size of 1 cm. The particle scattering cross sections are computed via $\sigma = \sum N * (a/2)^2 * \pi$ where $N$ is the number of particles and $a$ is the dust particle size.

Table 4
Spectral Energy Distribution Compiled from the Literature

| Wavelength ($\mu$m) | Flux (mJy) | Uncertainty (mJy) | Instrument | References |
|---|---|---|---|---|
| 2.1598 | 145530.00 | 1455.00 | this work | NA |
| 3.55 | 62860.00 | 250.00 | irac | Fazio et al. (2004) |
| 4.49 | 37150.02 | 150.00 | irac | Fazio et al. (2004) |
| 5.66 | 25370.00 | 350.00 | irac | Fazio et al. (2004) |
| 7.87 | 14450.00 | 70.00 | irac | Fazio et al. (2004) |
| 12.00 | 6688.80 | 535.10 | cciras | Neugebauer et al. (1984) |
| 23.67* | 2099.08 | 62.97 | mips | Rieke et al. (2004) |
| 25.00 | 1894.18 | 284.13 | cciras | Neugebauer et al. (1984) |
| 34.80* | 1300.00 | 90.00 | SOFIA | Su et al. (2017) |
| 60.00 | 1702.70 | 255.40 | cciras | Neugebauer et al. (1984) |
| 70.00* | 1739.50 | 10.00 | herschel | Greaves et al. (2014) |
| 71.42 | 1809.99 | 207.00 | mips | Rieke et al. (2004) |
| 100.00 | 1989.00 | 298.35 | cciras | Neugebauer et al. (1984) |
| 155.89 | 970.05 | 186.01 | mips | Rieke et al. (2004) |
| 160.00* | 1237.32 | 30.00 | herschel | Greaves et al. (2014) |
| 250.00* | 525.00 | 30.00 | submm | Greaves et al. (2014) |
| 350.00* | 268.00 | 20.00 | submm | Greaves et al. (2014) |
| 450.00 | 256.00 | 20.00 | JCMT/SCUBA-2 | Holland et al. (2013) |
| 500.00* | 164.00 | 10.00 | submm | Greaves et al. (2014) |
| 850.00* | 38.70 | 2.50 | JCMT/SCUBA | Greaves et al. (1998) |
| 1200.00* | 17.30 | 3.50 | submm | Lestrade & Thilliez (2015) |
| 1300.00* | 17.00 | 5.00 | ALMA | MacGregor et al. (2015) |

**Note.** Wavelengths marked with an asterisk (*) are included in the fit to the outer disk SED. For the data at 23.7 and 34.8 $\mu$m, we have removed the inner disk contributions given in Table 2. At 34.8 $\mu$m, we used the encircled energy data from the SOFIA measurements to estimate 310 mJy from the outer ring. We used the model from Debris Disk Simulator to estimate the contribution of the inner region to the total, estimating it to be ∼14% at 70 $\mu$m, and assumed the same fraction at the longer wavelengths. At this level, it has little effect on the fit to the SED.

In this way, a library of SEDs was generated for different radial locations of the disk that could be adjusted for changes in the dust particle size distributions, and scaling factors could be applied to each model SED to determine which disk radii are likely to contribute most strongly to the observations. To explore this degenerate parameter space quickly and efficiently, we employ an MCMC approach using the emcee package (Foreman-Mackey et al. 2013) and the Affine Invariant sampler (Goodman & Weare 2010). We rely on a $\chi^2$-based log-likelihood estimation given by Equation (2) to inform the MCMC.

$$\ln[P(D|\Theta)] = -\frac{1}{2}\left[\chi^2 + \sum_{i=1}^{N}\ln(\sigma_i^2) + N\ln(2\pi)\right]. \quad (2)$$

Here, $D$ and $\Theta$ represent the data and model, respectively, $N$ is the number of data points, and $\sigma$ is our uncertainty. The model $\Theta$ is generated from a linear combination of the components in our model SED libraries with scale factors as free parameters. These scaling factors are evenly spaced logarithmically with values from $10^{-6} - 1$. This effectively employs the resultant radial dust distribution in the dynamical model as a prior. Note, however, that the particle surface densities shown in Figure 10 only vary by a factor of $<10^3$ across all radii (half of the allowed parameter range), and this prior does not have a strong impact on the final MCMC results. We compare the results of this dynamical model with the MCMC results in Section 4.3.

Separate MCMC runs were performed to optimize for the inner+intermediate zone and the outer ring, respectively. The reasons for this were twofold: (1) the IRS slit widths (4.″7 and 11.″1 for the SH and LH modules, respectively) do not provide information on dust outside of the intermediate zone; and (2) as discussed above, initial testing required different dust properties to fit the IRS spectrum and the millimeter slope of the broadband SED arising from the outer ring. Accordingly, we conducted two MCMC runs, one optimized to fit the IRS spectrum, which is sensitive to material in the inner ∼5″ (radius) of the disk and the second optimized to fit the broadband SED and the outer disk. The broadband SED was corrected to only include flux from outside the inner region of the disk by subtracting the contributions from the inner regions using the information in Section 2.2 (see Table 4). These observables were tested against libraries of synthetic model spectra and SEDs drawn from the dynamical model, separated in radius, and coarsely sampling the particle size distributions. We elected not to force an arbitrary cutoff between the intermediate and outer disk for our two MCMC runs. In radial zones where the observables are agnostic to disk material (e.g., the outer, cold belt for the IRS spectra), we anticipate only upper limits for dust masses and surface densities.

The generation of spectral/SED libraries for different particle size distributions is computationally prohibitive, and only a coarse grid was explored. For both MCMC runs, we explored minimum grain sizes of $a_{\min}=0.25$, 0.5, 1.0, and 2.0 $\mu$m, i.e., from typical blowout sizes for spherical solid grains to nearly an order of magnitude larger, and assumed Mie scattering with the appropriate angular dependence. The MCMC run optimized for the IRS spectrum assumed a size distribution of power-law slope $p=3.8$. The MCMC run optimized for the outer disk SED left the power-law slope ($p$) as a free parameter ranging from 3.4–3.8, to include both the steep slope indicated for the intermediate zone in our reconnaissance and the range of slopes for the outer ring from the long-wavelength SED slope. Within the MCMC, values for the slope ($p$) and $a_{\min}$ were generated from a continuous distribution and then binned to the discrete library values before computing the likelihood. The aim was to roughly match the slopes of the IRS spectrum and SED, while suppressing the scattered light sufficiently to make the dust undetectable in the HST/STIS data set. A full exploration of dust properties (compositions, porosities, etc.) is left for future work.

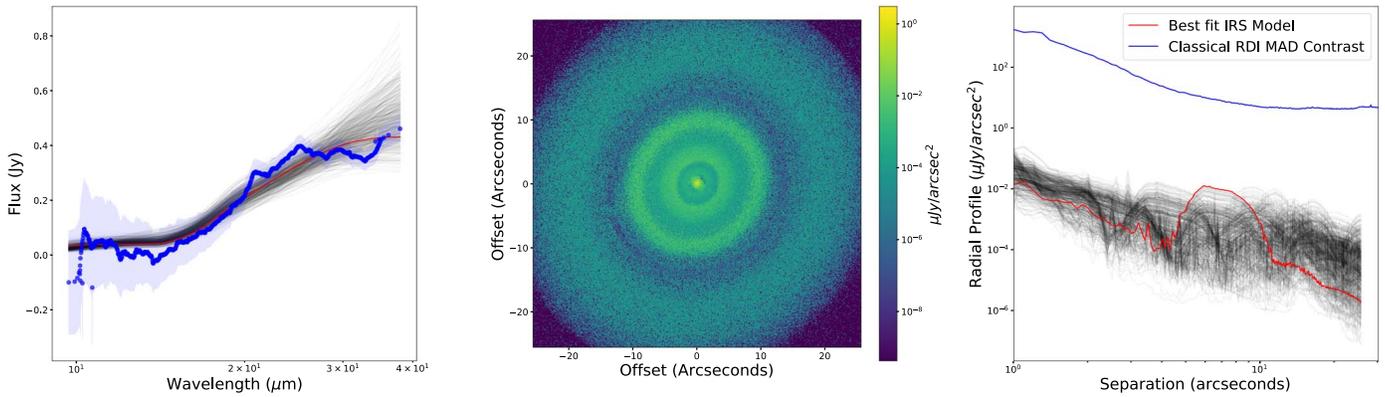

**Figure 11.** Scattered light prediction from a fit to the IRS spectrum. Left: the IRS spectrum compared to 500 models randomly drawn from the MCMC chain. Center: a synthetic image in scattered light based on the best-fit model to the IRS spectrum and generated at the STIS coronagraphic wavelength. Right: the radial profile in scattered light for the synthetic image (center). Radial profiles for the 500 alternative models from the IRS spectrum fits (left) are also shown. With a minimum grain size of 0.5–1 $\mu$m, scattered light remains well below our STIS detection limit.

Each run employed 200 walkers and 1,000,000 iterations with a 10,000 iteration burn-in. We discuss the results in Sections 4.2.3 and 4.2.4, with supplementary materials presented in Appendix C.

### 4.2.3. The IRS Spectrum Optimized MCMC Results

The parameter values for the model that provided the best fit to the IRS spectrum and the inner and intermediate disk are provided in Table 3. Full posterior distributions are presented in Figure 20, and the observed IRS Spectrum is compared to the best-fit model in Figure 11. This parameter space is highly degenerate, but we are able to rule out some values for the scale factors (which dictate the dust masses and surface densities), and place some constraints on the dust population.

As a test of convergence, we compute the effective sample size (ESS), which provides a measure of the effective number of independent samples in a correlated chain. This is computed via $\mathrm{ESS} = N_{\mathrm{samples}}/(2\tau_x)$ where $\tau_x$ is the integrated autocorrelation time for each free parameter $x$. The total number of samples $N_{\mathrm{samples}} = 10,000,000$. We compute the autocorrelation time via the method described in Goodman & Weare (2010) and find values ranging from 7300–10,000 with corresponding ESS values of $\approx$500–700 for all parameters. The standard error for a correlated MCMC can be expressed as $\sigma_n/\sqrt{N_{\mathrm{eff}}}$.

The IRS spectrum places the strongest upper limit for the disk at the 0–3 au bin with a best-fit dust mass of $\sim 10^{-7}\,M_\oplus$, while the narrow 3–4 au zone has the lowest dust mass in the best-fit model. Our scattered light observations are not able to place a constraint on the dust content in the 0–4 au region, but it is clear that some small amount of dust is required here to fit the IRS spectrum. This result is confirmed by the LBTI detection at $\sim 10\,\mu$m of a circumstellar component in this zone (Ertel et al. 2020).

The posterior distributions for dust in the intermediate disk (interior to 20 au) all have clear peaks near the largest allowed values with flat tails toward the lowest allowed values. The posterior distributions in this region are also L-shaped, meaning that a population of dust is required here, but the MCMC has no strong preference for the exact radial location. This is illustrated in the right panel of Figure 11, which shows the radial surface brightness profile in scattered light (a proxy for the dust scattering surface area) for the 500 best-fit models. As an example, the best-fit model scales the 12–20 au region to the lowest possible value of $10^{-6}$ (a dust mass of $2.7 \times 10^{-8}\,M_\oplus$; though higher values are also consistent with the observations). However, there is a peak in the dust density in the next zone, 20–30 au, in both the IRS and SED models. The posterior distributions for zones $>$37 au are flat with the largest dust masses excluded. This is unsurprising as the IRS spectrum is not expected to be sensitive to disk material at these separations.

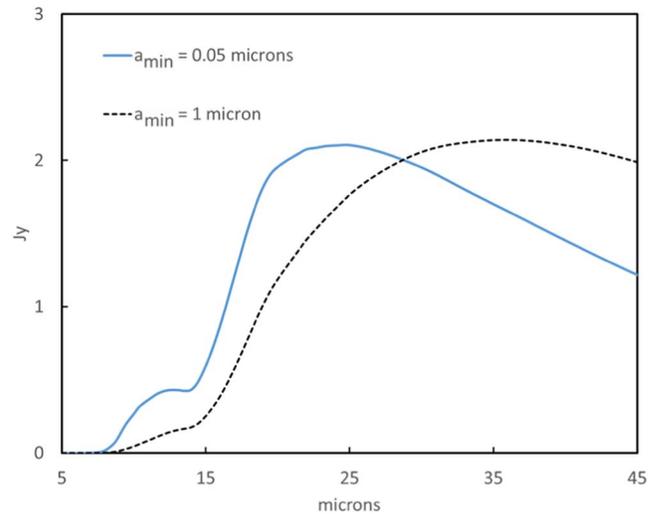

**Figure 12.** Comparison of SEDs for grain size distributions with slopes of $-3.8$ and minimum sizes of 0.05 and 1 $\mu$m. The normalization of the $y$-axis is arbitrary.

The IRS data set is fitted with nearly equal likelihood for both $a_{\min} = 0.5$ and 1 $\mu$m. A similar result, $a_{\min} = 1\,\mu$m, is also strongly preferred in the models by Su et al. (2017). The underlying reason is illustrated in Figure 12, for the value of $a_{\min} = 0.05\,\mu$m suggested by Reidemeister et al. (2011) and for 1 $\mu$m. Significant numbers of very small grains emit strongly in the 10–17 $\mu$m range and are not consistent with the lack of observed flux there, nor does a size spectrum extending to tiny grains emit strongly enough beyond 25 $\mu$m to match the data. The preferred values for $a_{min}$ are both larger than expected for the blowout size of spherical Mie grains for the dust composition used in our radiative transfer modeling ($\sim$0.3 $\mu$m; see Section 4.2.1). This could explain the low scattered surface brightness of $\epsilon$ Eri, and it is described in more detail in Section 5.1.

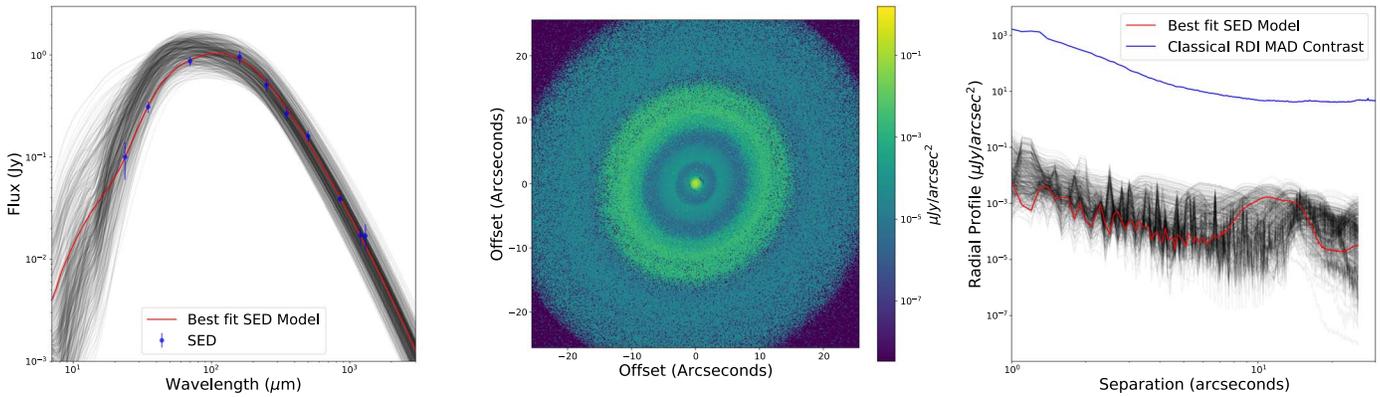

**Figure 13.** Scattered light prediction from a fit to the SED (outer ring). Left: the broadband thermal SED compared to 500 models randomly drawn from the MCMC chain. Center: a synthetic image in scattered light based on the best-fit model to the SED and generated at the STIS coronagraphic wavelength. Right: the radial profile in scattered light for the synthetic image. Radial profiles for the 500 models from the SED fits (left) are also shown. Again, with the minimum grain sizes required from fitting the thermal SED, the scattered light falls well below the detection limit.

We use the same model parameters to construct a synthetic image from the resulting dust distribution and compare the expected surface brightness to our achieved STIS coronagraphic image sensitivities (Figure 11, middle panel). The image shows a series of distinct rings in the inner regions of the disk. The radial profiles for a large representative set of models that can fit the IRS spectrum (Figure 11, right panel) show that the surface brightness is at least 2 orders of magnitude below our STIS surface brightness sensitivity. The use of Mie scattering may artificially minimize the signal at this large scattering angle in our models. However, calculations for spheroidal particles (e.g., Lee 2013) and laboratory measurements of mineral dust particles sized similarly to the dust in the IRS zone (e.g., West et al. 1997) show the effects to be modest, i.e., a factor of $\sim$3, and insufficient to lift the predicted flux into the detectable range. That is, because of both the large minimum grain size required to fit the thermal spectrum (we exclude the smallest grains with high scattering efficiencies), the large scattering angle, and also a more radially extended intermediate disk, the dust content of our models is sufficiently low to be undetectable in scattered light.

*4.2.4. The Outer Disk Broadband SED Optimized MCMC Results*

Now we present the results for the MCMC run optimized for the outer disk using broadband photometry compiled from the literature. The modeled best-fit values are summarized in Table 3. Full posterior distributions are presented in Figure 21, and the broadband SED (corrected to include only flux from the outer, $r > 10''$ disk) is compared to the best-fit model in Figure 13. We use the effective sample size to test the convergence as described above. Autocorrelation times were shorter in this case and ranged from 2000–2500 with corresponding effective sample sizes of 2000–2500 for all parameters.

The posterior distributions for all zones from 0–20 au are flat at low values and drop off at scale factors $\lesssim 10^{-2}$. We did not expect the long-wavelength broadband SED to place a constraint on dust in these innermost regions. These flat posteriors do not rule out the values preferred by the IRS optimized MCMC, and we defer to the IRS best-fit model interior to 20 au.

Moving outward, as is the case in the IRS models, the 20–30 au and 30–37 au zones allow for higher dust masses (excluding only scale factor values $> 10^{-1.5}$) with values of $1 \times 10^{-5}$ and $5 \times 10^{-4}$ $M_\oplus$, respectively. Both the SED and IRS models show a substantially reduced density in the 37–44 au zone, coincident with a clearing of material by $\epsilon$ Eri c in the dynamical models.

Outside of 40 au, the posterior distributions for the outer disk optimized MCMC all have peaks near the largest allowed values with long tails to the lower end of the allowed range and have the characteristic L-shape, illustrating the degeneracies among these zones. The right panel of Figure 13 demonstrates the diversity of dust locations within well-fitting models to the observations, which are consistent with the belt imaged by ALMA. Our best-fit model peaks in the 44–52 au zone with a dust mass of $4 \times 10^{-3}$ $M_\oplus$. This peak at the location of $\epsilon$ Eri c where larger particles are trapped in a 1:1 resonance, combined with the lack of flux just interior, agrees well with the Booth et al. (2017) predictions.

The minimum grain size is best fit by the largest allowed value of 2.0 $\mu$m, which we treat as a lower limit. This is larger than the $a_{min}$ value preferred by the IRS model and is significantly larger than the blowout size for any reasonable combination of dust particle compositions, porosities and/or geometries (see discussion Section 5.1). The $a_{min} \geqslant 2$ $\mu$m value is sufficient to significantly suppress the scattered light surface brightness for the disk despite the much larger total disk mass than the model preferred by the IRS spectrum. Unlike the IRS spectrum optimized MCMC run, the SED optimized MCMC run prefers the shallowest grain size distribution slope of the allowed parameter space with a peak at $p = 3.4$. This will weight the particle size distribution more heavily toward larger grains. These results suggest that the grain size distribution for the $\epsilon$ Eri system is radially dependent with the inner disk comprising mainly warm, small dust particles and the outer disk dominated by the primordial planetesimal belt.

We use the same model parameters to construct a synthetic image from the resulting dust distribution and compare the expected surface brightness to our achieved STIS coronagraphic image sensitivities in the right panels of Figure 13. The outer ring is faintly visible in the simulated images, but the 2.0 $\mu$m minimum grain size lowers the scattered light surface brightness almost 3 orders of magnitude below our STIS surface brightness sensitivity limit. The Proffitt et al. (2004) direct subtraction results are deeper than ours at the outer ring, but the predicted scattered light is still 2 orders of magnitude below them.

*4.2.5. Supplemental MCMC Optimized for the Complete SED*

Detailed information about the infrared spectrum and location of the disk components is not available for the vast majority of debris disks. Therefore, we carried out a third MCMC run that was constrained only by the photometric points (including contributions from the inner disk), as an experiment to see if the attributes we describe as unique for the $\epsilon$ Eri system might be present in other cases but not revealed by the existing data. These results are presented in full in Appendix C. In summary, the scale factors agree with the outer disk corrected SED MCMC results presented above with the exception of the 0–3 au region, which is tightly constrained. These data also prefer a minimum grain size of 2 $\mu$m, while the size distribution parameter $p$ is broadly peaked at 3.5 representing a compromise between the IRS spectrum and outer disk SED results above. This provides further evidence for a radially dependent grain size distribution.

*4.3. Comparison with Dynamical Modeling*

The results from the SED modeling in Table 3 can be compared with the zones where debris disks are likely from the dynamical model illustrated in Figure 10. The radii derived in the SED model are uncertain, since modest adjustments in grain optical properties would change them. However, with this proviso, the photometric modeling is very consistent with the dynamical model. The specific model in the table suggests that the location of the intermediate belt is near the middle of the permitted zone of 4–40 au, but a range of positions within that zone is permitted within the uncertainties. The model is consistent with the presence of a gap where $\epsilon$ Eri c is hypothesized to lie, and indicates a broad zone of infalling dust inside the outer ring detected by ALMA. This latter structure is consistent with the models of PR drag on dust released from the outer ring as derived by Reidemeister et al. (2011); the combined dynamical and photometric modeling suggests that this inflow is interrupted by $\epsilon$ Eri c.

## 5. Discussion

In the following sections, we place these modeling results in context. These fits to both the IRS spectrum and the broadband SED of the outer disk require a large minimum grain size and a radially stratified particle size distribution. The dust distributions are generally broad, and while narrow rings of material are not ruled out, there is no strong preference for any particular radial location. In general, all best-fit models push the majority of the dust mass to wider separations than Su et al. (2017) with less incident flux. This combination of larger grain sizes (either from porous grains or a suppression of small grain production) and broad radial dust distributions (corresponding to a decrease in the disk surface brightness) significantly suppresses the scattered light flux. In Section 5.1 we discuss the dust opacity in more detail, and in Section 5.2 we examine the impact of forthcoming JWST observations of this system.

*5.1. Dust Opacities*

We have found that the very faint scattered light signal from the $\epsilon$ Eri debris disk is explained by two dust scattering effects: (1) the face-on orientation of the system; and (2) the large minimum grain size.

The near-face-on orientation of the $\epsilon$ Eri disk results in a minimum scattering angle probed by our data set of 60°, so we miss the bright forward scattering peak of the emission. Dust particles only scatter efficiently when the dimensionless size parameter is smaller than the wavelength, i.e., when $2\pi a \simeq \lambda$. For STIS ($\lambda_c \sim 0.58$ $\mu$m), this occurs for particle sizes of $a \simeq 0.03$–$0.16$ $\mu$m over the wide STIS bandpass. Grains significantly larger than these values have strongly anisotropic scattering that is strongly suppressed at large angles. Therefore, the large minimum grain size preferred by both MCMC runs, $a_{\min} = 0.5$–$1$ $\mu$m for the intermediate disk and $\geqslant 2$ $\mu$m preferred by the SED model, also helps explain why the dust is not seen in scattered light with HST/STIS.

Our modeling based on Mie scattering by spherical grains represents the most extreme example of anisotropic scattering. It is possible that more realistic grain properties, such as aggregates, would result in less anisotropy. However, the size of the difference between the Mie-scattering predictions and the detection limits indicates that the scattering anisotropy is a major contributor to the low signal level.

Our exploration of the dust particle property parameter space was very sparse. For both the outer disk optimized and full SED models, the posterior distribution for the minimum grain size was pushed up to the largest allowed values, and extending toward larger values may have improved the fits. Thus, we do not claim that the minimum grain size is necessarily $\sim$1–2 $\mu$m, but that the grains have a color-dependent opacity that mimics spherical Mie grains with the minimum grain size preferred by the models.

In either case, the minimum particle size preferred by our models is larger than the blowout size expected for spherical Mie grains (see Section 4.2.1). There are two classes of possible explanation. First, for the $\epsilon$ Eri system, a blowout size $\sim$1 $\mu$m is possible assuming highly porous grains or more complex agglomerates. Second, the production of very small grains may be substantially reduced due to dynamical effects around low-mass, low-luminosity stars. Pawellek & Krivov (2015) found that the ratio of the minimum grain size to the blowout size increases with decreasing stellar luminosity, and is $\sim$10 at $L_\odot$ (they did not study stars of lower luminosity). Either explanation would contribute to a lower dust scattering albedo. Chen et al. (2020) recently found that a collection of large, irregular grains (consistent with Discrete Dipole Approximation calculations) was necessary to fit multiwavelength scattering phase functions to the A star debris disk host, HR 4796A. It is likely that a combination of more complex prescriptions for dust particles and radiative processes will be needed to reproduce the next generation of debris disk observations.

Evidently, the size distribution power-law slope, $p$ varies radially. The IRS spectrum prefers a very steep grain size distribution with $p = 3.8$, while the SED optimized for the outer disk prefers the lowest allowed value of $p = 3.4$, and the full SED (see Appendices A, C) provides a compromise of $p \sim 3.65$. Combined with the $a_{\min}$ results, it is clear that the IRS spectrum requires a population of small dust grains in the inner disk, while suppressing the largest grains. Conversely, the long-wavelength SED requires a population of larger dust grains in the outer disk in line with the outer ring imaged in the millimeter. This difference in grain populations is compelling and persists regardless of radial dust distributions, despite large degeneracies between the belt locations and grain properties (which were sampled sparsely). The smaller $a_{\min}$ value preferred by the IRS spectrum probing the inner 5″ of the disk may also point toward a radially dependent minimum particle size. We leave a complete exploration of radially varying dust populations for future work.

### 5.2. Epsilon Eridani and JWST

Two JWST Cycle 1 GTO programs will be observing $\epsilon$ Eri, one with the Mid-Infrared Instrument (MIRI) and the other with the Near-Infrared Camera (NIRCam). The MIRI program will resolve the thermal flux of the Asteroid Belt–component using the 15.5 $\mu$m four-quadrant phase mask (4QPM) coronagraph. The 23 $\mu$m Lyot coronagraph and 25.5 $\mu$m imaging will resolve the entire system. The MIRI images will highlight the location of the larger dust particles ($\leqslant 10$ $\mu$m), which are less affected by radiative forces and therefore closely track the location of the parent body belts. Observations with the NIRCam coronagraph will be used to search for planets, including Epsilon Eridani b, with a wide filter at 4.44 $\mu$m and with control observations at 3.22 $\mu$m. A direct imaging detection of $\epsilon$ Eri b would allow for much tighter constraints on its orbital parameters and, hence, its degree of interaction with the surrounding disk. Additionally, we find evidence for $>1$ $\mu$m dust locked in a 1:1 resonance with $\epsilon$ Eri c. An asymmetry in the disk observed with JWST at that location would help to confirm the presence of the planet. Spatially locating planets and parent body belts interior to the Kuiper Belt analog would help to elucidate the nature of the intermediate dust content.

### 6. Summary and Conclusions

We report on a deep, but unsuccessful, search for scattered light from the $\epsilon$ Eri debris system using HST. We have combined the resulting limits on scattering with results from the literature in an MCMC framework that provides robust constraints on the dust size distribution and a more complete picture of the radial locations where dust is permitted. The result is a revised picture of the $\epsilon$ Eri system architecture.

The disk likely has a population of small dust grains in the very inner regions of the disk, interacting with the known planet $\epsilon$ Eri b and in part detected interferometrically (Ertel et al. 2020). The IRS spectrum (see Section 4.2.3 and Appendix C) has a strong preference for an intermediate disk located in the 6–37 au region, with some preference to lie in the outer half of this zone. The outer ring at $\sim 69$ au dominates the broadband far-infrared SED. The modeling of this component permits a broad distribution of dust between the orbit of the putative planet $\epsilon$ Eri c (at 44 au) and the ring, suggestive of dust flowing inward from the ring due to PR drag, but does not exclude a single, cold belt.

The minimum grain size preferred by all of our SED modeling of the outer component (outside of 40 au) was large ($\geqslant 2$ $\mu$m), while the $a_{\min}$ value preferred for the inner component (within 40 au) by the IRS spectrum was smaller ($\sim 0.5$–1 $\mu$m), though still larger than that expected for the blowout size predicted for our dust population. This tension could be reduced by invoking more complex grain geometries (e.g., agglomerates and high porosities) or via a reduction in the small grain production due to dynamical effects around this low-mass star (see Pawellek & Krivov 2015).

The grain size distribution power-law slope also appears to vary with radius. The IRS spectrum preferred the steepest allowable slope ($-3.8$) for the inner and intermediate disk regions region (0–44 au), while the outer disk ($\sim 70$ au) SED preferred the shallowest allowed slope ($-3.4$). This implies that the inner regions of the disk are more heavily weighted toward the smallest grains, while the outer disk is weighted toward the larger ones.

There is a long history of the scattered light being too faint to match models of the thermal output of debris disks (e.g., Krist et al. 2010; Golimowski et al. 2011; Lebreton et al. 2012). We show that the probable cause, at least for this star, is the assumption that the grain sizes extend down to the blowout size when in fact, the grains are significantly larger. Upcoming JWST coronographic observations will pinpoint the locations of the dust belts and provide radially resolved color information to further constrain the dust particle properties. $\epsilon$ Eri remains one of the most powerful systems to test models of circumstellar disk evolution.

We thank G. Schneider for sharing his vast knowledge on classical PSF subtraction of HST/STIS observations and K. Su for discussions on $\epsilon$ Eridani disk models. We also thank B. Ren for consultations on nonnegative matrix factorization and more generally on the efficacy of various post-processing algorithms given the noise properties (incredible surface brightness sensitivity) of HST/STIS coronographic observations. This research is based on observations made with the NASA/ESA *Hubble Space Telescope* obtained from the Space Telescope Science Institute, which is operated by the Association of Universities for Research in Astronomy, Inc., under NASA contract NAS 5–26555. These observations are associated with program 15906.

*Facility:* HST (STIS).

*Software:* astroconda, autofillet http://stis2.sese.asu.edu/code/, DiskDyn https://github.com/merope82/DiskDyn, pyKLIP https://bitbucket.org/pyKLIP/pyklip/src/master/, NMF https://github.com/seawander/nmf_imaging.

### Appendix A
### Spectral Energy Distribution

We present a spectral energy distribution (SED) compiled from the literature for the $\epsilon$ Eri system. Here we have selected those measurements that contain information on the full spatial extent of the debris disk system and have removed literature photometric measurements that are limited to the innermost regions (see for example Table 2) and some millimeter measurements that only contain flux from the outermost debris ring. These values are used in the broadband photometry MCMC modeling discussed in Appendix C below.

## Appendix B
## Details of the Coronagraphic Reductions

### B.1. Classical Reference Differential Imaging (cRDI)

The primary objective during classical RDI post-processing is to coadd as many PSF-subtracted target images as possible, without introducing unnecessary numerical/processing noise into the reduction. The order in which the processing steps are executed is crucial in obtaining the highest-contrast results. One may perform PSF subtractions on each individual exposure, finding the best-matching individual PSF reference image, and then translate these (shift and rotate) to a common coordinate system to combine. Another option is to translate the images first and then perform the subtraction. This is generally better, as the image translations fit higher-order functions to the data (we used a 3D sinc in IRAF, with a 9 px wide interpolation box). The PSF-subtracted images are not as smooth as the science images, and therefore additional noise will be introduced if they are translated instead of the science frames. An additional process to consider is to first combine some subset of the science images, if they were taken in an uninterrupted sequence and with tracking ensuring no PSF broadening. This helps, as it results in an even smoother image prior to translation.

We first examined the positioning of $\epsilon$ Eri in our images, to see whether the science frames could be combined prior to translations. In Figure 14, we show the location of $\epsilon$ Eridani in image frame coordinates for the first epoch of observations at both coronagraphic apertures. These positions were determined by the "X marks the spot" method, where the OTA diffraction spikes are fitted with two lines, and their intersection defines the location of the host star. According to the HST DrizzlePac Handbook (Gonzaga et al. 2012), the tracking with fine lock on two guidestars—which was used in our program—has a typical rms accuracy of 2–5 mas (0.04–0.1 px) or less throughout an orbit. That means that the majority of the 0.25 px dispersion in positions we see is from our "by-eye" fitting, rather than tracking issues and that the tracking keeps the target within 10% of the stellar PSF's FWHM. If the positions did have a 0.25 px dispersion, they would still be within 20% of the stellar PSF's FWHM. This high accuracy in tracking allows us to first combine all images taken within the same orbit at the same coronagraphic aperture. We generated combinations with median and mean averaging, using data with $1\sigma$, $2\sigma$, and $3\sigma$ clipping around the data median value for all visits, including the PSF reference observations.

The HST OTA experiences thermal expansion over each orbit, known as "breathing" (Hasan & Bely 1994), which results in slight variations to its PSF. These variations will affect the radial profile of the occulted central source and therefore the coronagraphic contrast. The level of breathing is not predictable, but is something that generally settles over time as long as the observatory stares in the same direction. We examined the effects of breathing on our data set, by plotting the ratios of the radial flux profiles of the PSF observations ($\delta$ Eridani) to the target observations of $\epsilon$ Eri. These ratio curves, plotted in Figure 15, show that the OTA settles over time and that the target observations in the last orbit at each epoch were the most similar to the PSF observations in the preceding orbit.

To counteract the effects of breathing, we scaled the observed target images acquired in the first two orbits at each epoch to the very last orbit. We did this by determining the best-fitting polynomial to the radial profile ratios, as plotted in Figure 16. These scaling functions were applied to the data prior to translation and PSF subtraction. Importantly, following PSF subtraction, the scaling functions were applied inversely, to scale back any existing extended and point-source fluxes to their actual observed value.

The linear translations between the target images and the PSF were determined by eye, using Image Display Paradigm #3 (IDP3), an IDL-based code that allows for easy visual tuning of scaling and shifts. Based on the visual fitting, the errors on our determined shifts are below 0.05 px. At 0.1 px offset from "best" centering, the subtraction residuals were clearly evident, allowing high-precision relative offsets to be determined between the target and corresponding PSFs.

Masks for each image were generated at their original reference frame coordinates and translated simultaneously with their corresponding image. The masks cover various imaging and detector artifacts, including: the coronagraphic mask of STIS, the OTA diffraction spikes, hot pixels and corresponding bleeding lines, the saturation bars we observed at WEDGEA2.5 in epoch 2, a bright background source near δ Eridani and its corresponding bleeding columns, and any apparent imaging artifact that we found in by-eye examination. These masks, unlike their corresponding images, were translated using only a bilinear formula to avoid ringing, which would be present if the sharp edge masks were translated with a bicubic (or higher-order) equation. Following translation, the mask values below 1 were rounded down to zero, thereby conservatively expanding the size of the mask by at most 1 px in each direction.

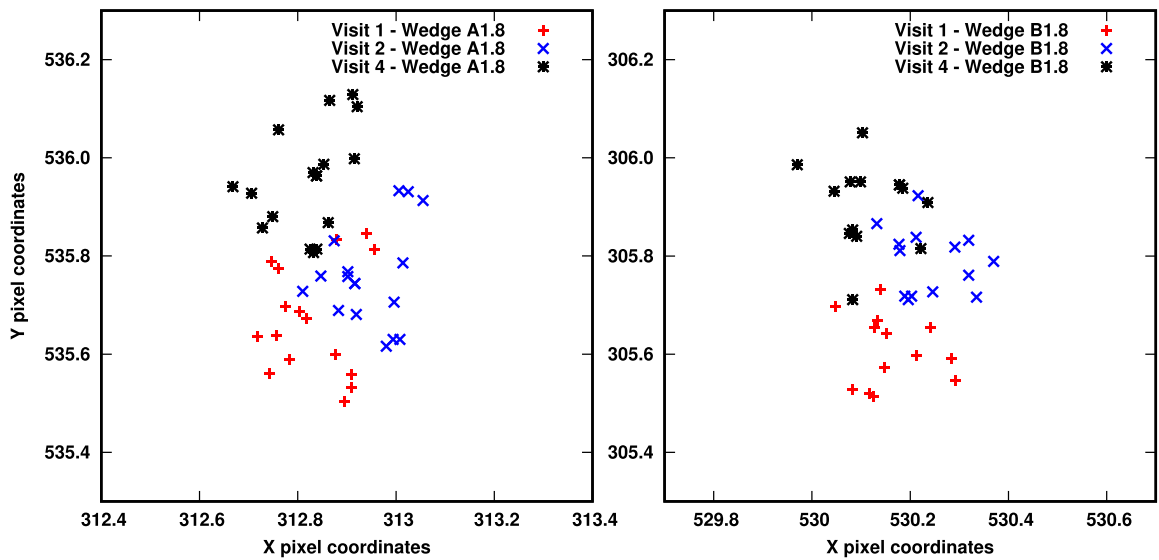

**Figure 14.** The measured position of $\epsilon$ Eri in visits 1, 2, and 4 during the first epoch of observations for each individual exposure. The location of the occulted star was determined by the fitting of the OTA diffraction spikes. The precision of this method is around 0.25 px in each direction, based on analysis of multiple previous HST/STIS data sets (e.g., Gaspar & Rieke 2020). As the standard deviation of each position is around this value per visit, the tracking of HST had to have errors much smaller. This conclusion allowed us to combine all images taken at each coronagraphic aperture per each visit, without worrying about PSF broadening.

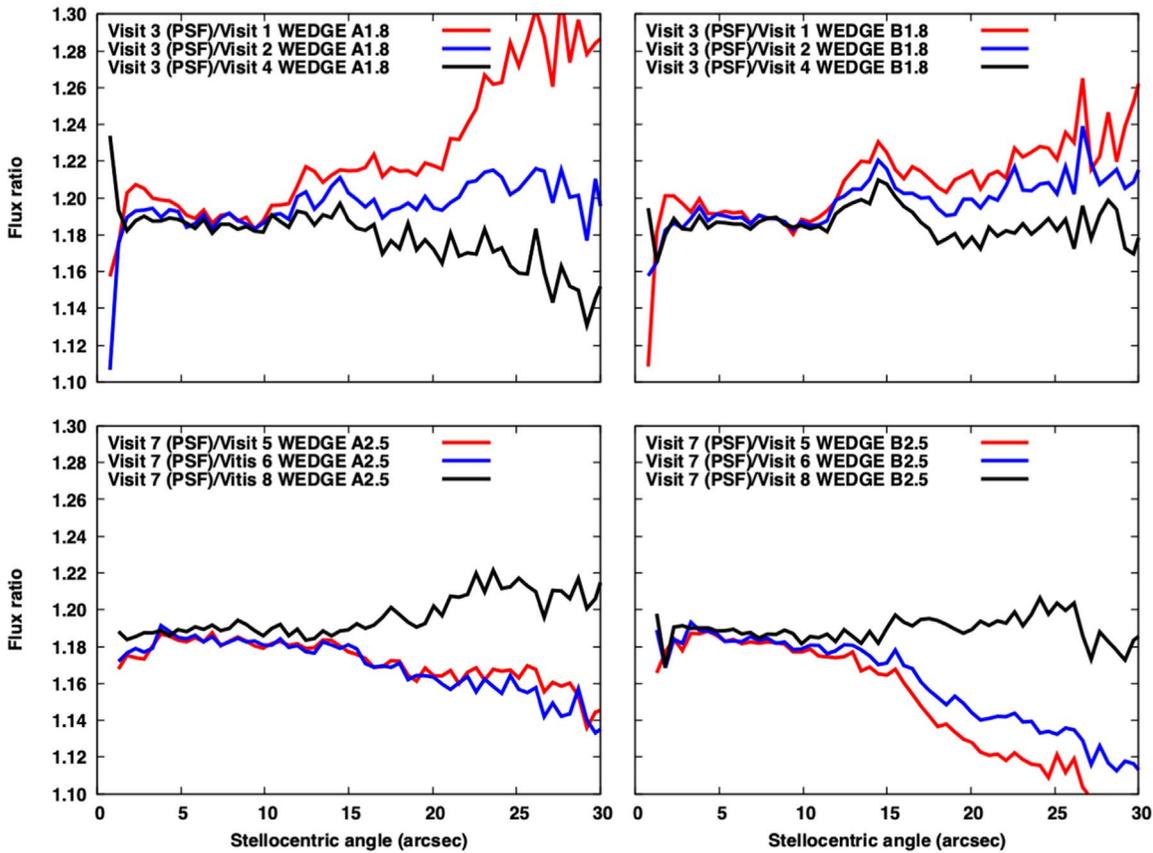

**Figure 15.** These figures show the ratios of the radial flux profiles of the PSF to ϵ Eri for all visits and coronagraphic apertures in our program. The plots highlight that the target observations themselves changed over the four orbit visit sets for each epoch, likely due to the "breathing" (thermal expansion) of the HST OTA. The telescope settles to a quasi-equilibrium by the third orbit within each set (visits 3 and 7; PSF observations), which is why the ratio function is mostly flat for the last orbit within each set (visits 4 and 8).

The locations of ϵ Eri were determined in the combined images (per orbit) using the tracking of the diffraction spikes method and together with their corresponding masks were translated onto a blank image large enough to frame all epochs at their aligned orientations. The two PSFs (one for each epoch of observations) were translated for each combined visit level image using the linear offsets determined with the "by-eye" fitting and the locations of ϵ Eri in the original images. The PSF observations were then subtracted from the target observations, and the subtracted images were re-scaled with the polynomials determined in the fitting shown in Figure 16. Finally, all target images were combined, using the rejection masks and weights based on the frame exposure times and the number of images included in the visit level combined image. As for the per-epoch combinations, we tried both mean and median combinations, using $1\sigma$, $2\sigma$, and $3\sigma$ clipping around the data median.

### B.2. Karhunen–Loéve Image Projection

We used the same set of base images as in the classical subtraction case wherein the images were processed through the calstis pipeline, video-noise removed, converted to units of counts per second, and a mask was applied to the wedge positions, OTA diffraction spikes, bad pixels, and bright point sources in the field. Unlike in the classical PSF subtraction case described above, here each individual exposure was treated as a unique frame by the subtraction algorithm; i.e., no observations were binned. This resulted in 171 individual exposures. The location of Epsilon Eridani behind the coronagraphic mask for each frame was determined automatically within pyKLIP using the OTA diffraction spikes and a Radon Transform algorithm to find the intersection point with a demonstrated precision of 0.1 pixel (Ren et al. 2019b). The determined coordinates are shown in Figure 17. These values were not identical to those found via the by-eye approach described for the classical PSF subtraction (see Figure 14) but the spread in x- and y-positions was equivalent both for a single visit (∼0.3 pixels) and for all visits at a single wedge position (∼0.5 pixels). The wider wedge positions contributed to slightly worse centering positions. In the case of WEDGE A2.5, this is likely due to interference from the saturated rows contributing additional uncertainty to the y-direction position for all visits. The WEDGE B2.5 data set has a very small spread within each visit, but the jumps between visits are larger along the y-direction (perpendicular to the wedge orientation), likely due to a pointing drift between subsequent visits.

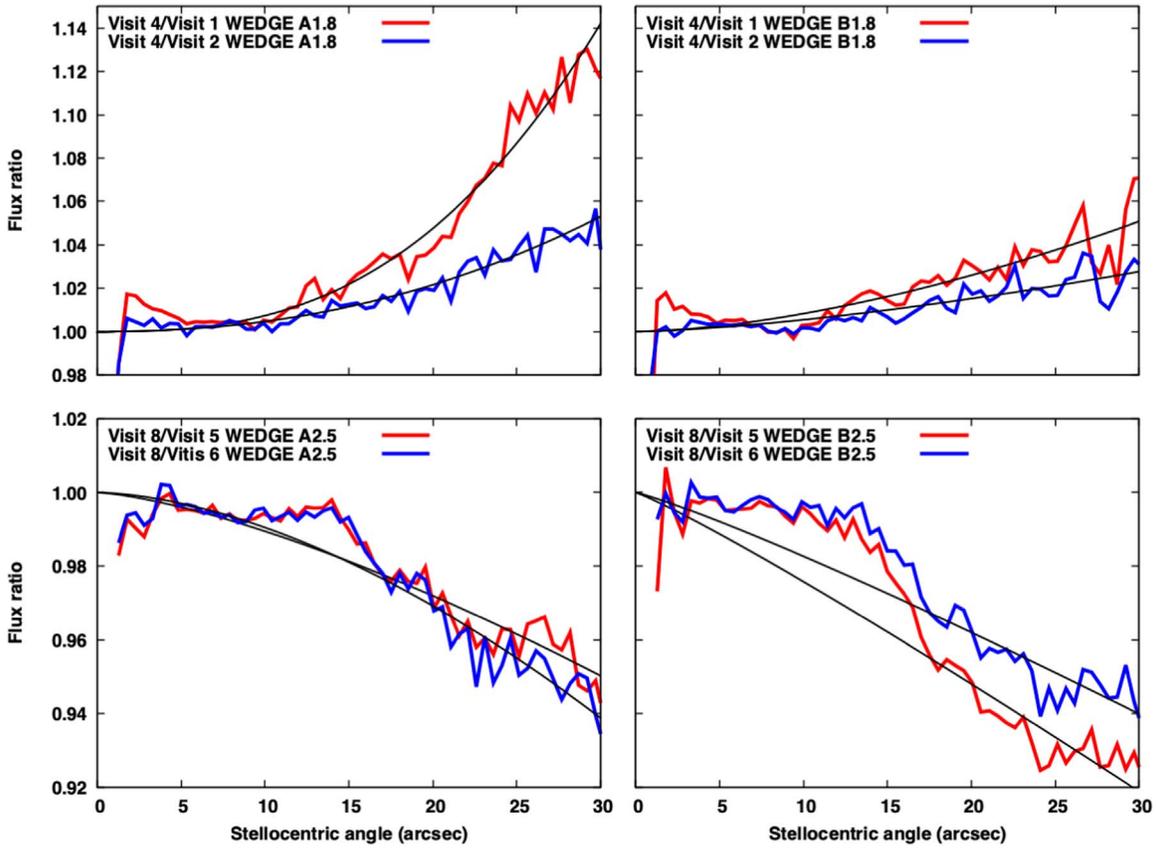

**Figure 16.** These figures show the ratios of the radial flux profiles of the last orbits within each set (target visits 4 and 8) to the first two target visits and the parabolic functions that were fitted to them, between 4″ and 30″ stellocentric angles. These functions were used to scale the target observations for PSF subtraction and then were inversely applied to correctly re-scale any possible extended features.

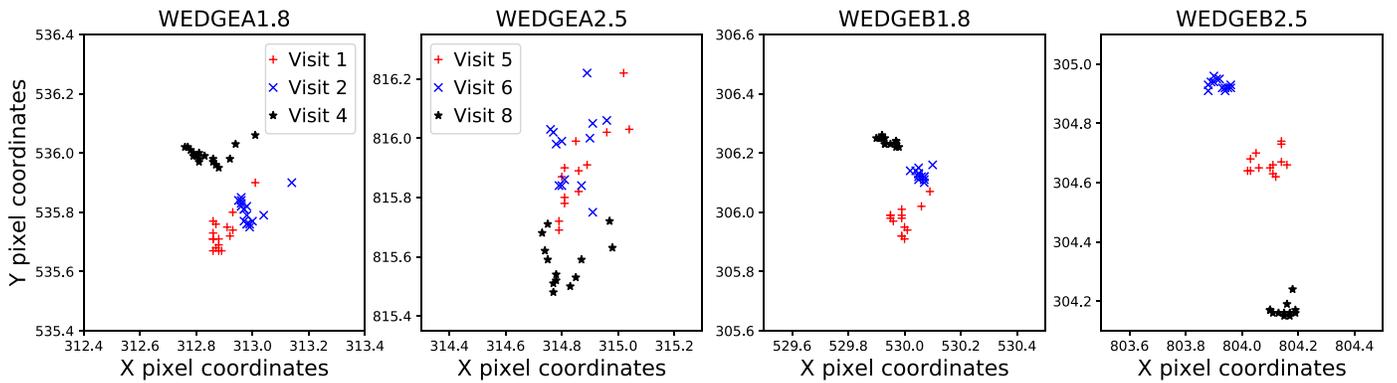

**Figure 17.** For both PCA PSF subtraction algorithms, the location of $\epsilon$ Eri behind the coronographic mask was determined using a Radon Transform centering algorithm and the OTA diffraction spikes. The pixel coordinates are shown here for all wedge positions and each of the 171 individual exposures. Unsurprisingly, the wider wedge positions, with the star placed closer to the edge of the detector, give slightly less accurate centers. Note that all axes have a range of 1 pixel.

The KLIP algorithm first assembles libraries for the Science and Reference Images, aligns and centers each exposure and subtracts the mean. Next, the KL transform of the set of reference PSFs is computed, and the five most correlated KL modes are chosen. Then it computes the best estimate of the target PSF from the projection of the target image on the KL eigenvectors, and finally the PSF-subtracted target image is calculated. pyKLIP also allows the user to specify individual zones within the frame to perform this background correction independently. For extended structures, it is customary to use a single zone, which we present here. We also performed KLIP using 20 radial zones or subannuli (evenly spaced linearly in bins of $\approx$2″). In the case of a face-on disk, any azimuthally symmetric disk material in the subannuli will be treated as background to be subtracted, and we would not expect to detect any disk signal. However, this does allow us to investigate the radial change in the slope between the science and target PSF. The results with and without the use of subannuli agree extremely well except for a small peak in the subannuli from 3″–4″. This is slightly exterior to the potential excess seen in the classical RDI results (see Figure 5) but could be indicative of a change in the relative PSF slopes around 2″.

### B.3. Nonnegative Matrix Factorization

The algorithm is described in detail in Ren et al. (2018). In short, NMF first decomposes the reference matrix into the product of two positive-value matrices ($R \simeq WH$) where $W$ is the coefficients matrix, and $H$ is a matrix of the components. In the case of sequential NMF, the first component is constructed and the corresponding coefficient and component matrices are used to initiate the construction for two additional components and so forth, resulting in a ranked component matrix. (Recall for the KLIP algorithm, the entire basis set is computed first and then the components are ranked by the magnitude of the eigenvalues before projection.) Once the basis set of NMF components has been constructed sequentially, the science target ($T$) images are modeled by minimizing $||T - \omega H||^2$ where $\omega$ is the coefficient matrix for the target. This is analogous to the least-squares approximation performed with KLIP, though values of $\omega$ tend to be smaller than the absolute values of the KLIP eigenvalues. We employ the NMF_IMAGING PYTHON package provided by Ren (2020).

The data were reduced identically to the classical RDI and KLIP subtractions: the images were first processed through the calstis pipeline, video-noise removed, converted to units of counts per second, and a mask was applied to the wedge positions, OTA diffraction spikes, bad pixels, and bright point sources in the field. Each of the 171 exposures was considered independently, and the location of the star behind the STIS coronagraphic wedge was determined using the Radon Transform algorithm as described for the KLIP PSF subtractions. Unlike with the KLIP reduction above, NMF was performed separately for each wedge position, and then all four (WEDGEA1.8, WEDGEA2.5, WEDGEB1.8, and WEDGEB2.5) were median combined.

## Appendix C
## Supplemental MCMC Results

We conducted a separate MCMC run using the uncorrected broadband photometry at all available wavelengths (see Figures 18, 19) to supplement the two MCMC runs optimized for the IRS spectrum (Figure 20) and outer disk photometry (Figure 21). This data set allows us to span the full radial range

**Table 5**
Dynamical and SED Modeling Results

| Zone (au) | Mass ($M_{Earth}$) | Surface (m$^2$) |
|---|---|---|
| 0–3 | 2.83e−06 | 3.79e+19 |
| 3–4 | 1.18e−08 | 2.70e+17 |
| 4–6 | 6.61e−08 | 1.67e+18 |
| 6–12 | 1.82e−07 | 4.82e+18 |
| 12–20 | 8.15e−08 | 2.19e+18 |
| 20–30 | 4.77e−08 | 1.28e+18 |
| 30–37 | 1.03e−07 | 2.74e+18 |
| 37–44 | 7.42e−06 | 1.89e+20 |
| 44–52 | 1.91e−06 | 5.08e+19 |
| 52–63 | 5.21e−06 | 1.41e+20 |
| 63–85 | 4.87e−03 | 1.33e+23 |
| 85–100 | 8.84e−07 | 2.43e+19 |
| 100–130 | 2.94e−07 | 1.11e+19 |
| Sum of Zones | 4.9e−03 | 1.3e+23 |
| Dust Particle Properties | | |
| Size Dist. (p) | 3.65 | |
| $a_{min}$ ($\mu$m) | 2.0 | |

**Note.** We present the parameter values for the best-fit model MCMC results optimized for the full literature-compiled SED. Mass and particle scattering cross sections for each radial zone (as described in Section 4.1) are computed from the best-fit scale factors. Masses are computed using the best-fit dust properties listed and a maximum particle size of 1 cm. The particle scattering cross sections are computed via $\sigma = \sum N * (a/2)^2 * \pi$ where $N$ is the number of particles and $a$ is the dust particle size.

present in our dynamical models, and is more directly comparable to SED-based debris disk modeling in the literature. The parameters of the MCMC were identical to the previous runs, with only the data set changing. The results are summarized in Table 5. Full posterior distributions are presented in Figure 19 and the literature-compiled broadband SED is compared to the best-fit model in Figure 18. We use the effective sample size to test the convergence as described above. Autocorrelation times ranged from 5000–7000 with corresponding effective sample sizes of 700–1000 for all parameters.

In many ways, the posterior distributions here match those seen in the MCMC run that was optimized for the outer disk broadband photometry. The 3–30 au regions all have flat posteriors falling off toward higher values. A peak to the posterior at higher values begins to emerge in the 30–37 au zone and becomes stronger moving outward. Some strong dust contributions are required to fit the SED in the 52–100 au regions, colocated with the sharp ring detected with ALMA, though dust is not required in all three zones in that range. Finally, the posterior distribution for the extended halo zone outside of 100 au is flat, and the SED is impartial to dust in this region.

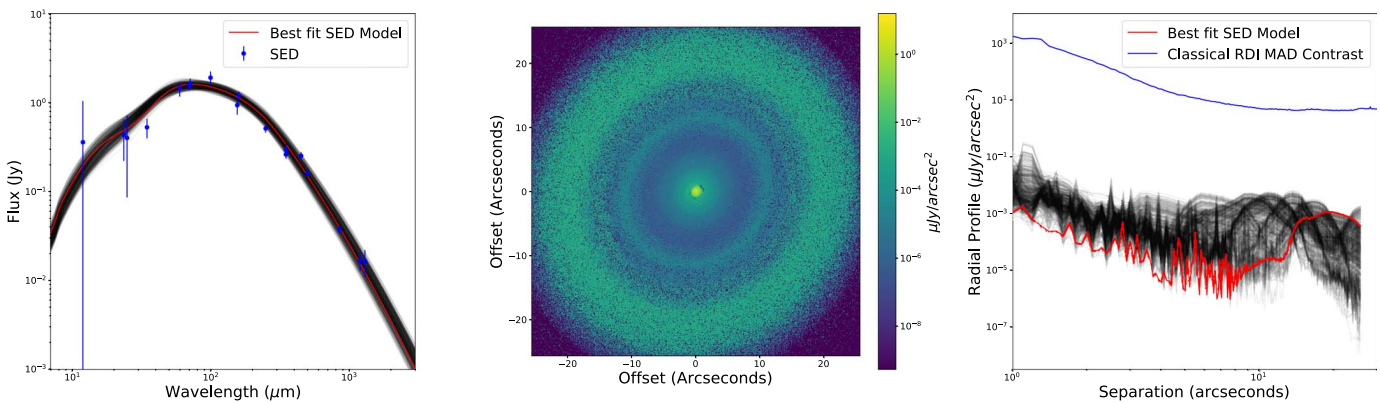

**Figure 18.** The best-fit model for the broadband optimized MCMC. Left: the broadband SED is compared to the best-fit model (red) and 500 models randomly drawn from the MCMC. Center: a synthetic image generated at the STIS coronagraphic wavelength for the best-fit model. Right: the radial profile for the synthetic image is shown against the image sensitivity of our HST observations. Radial profiles for the 500 models from the SED figure (left) are also shown.

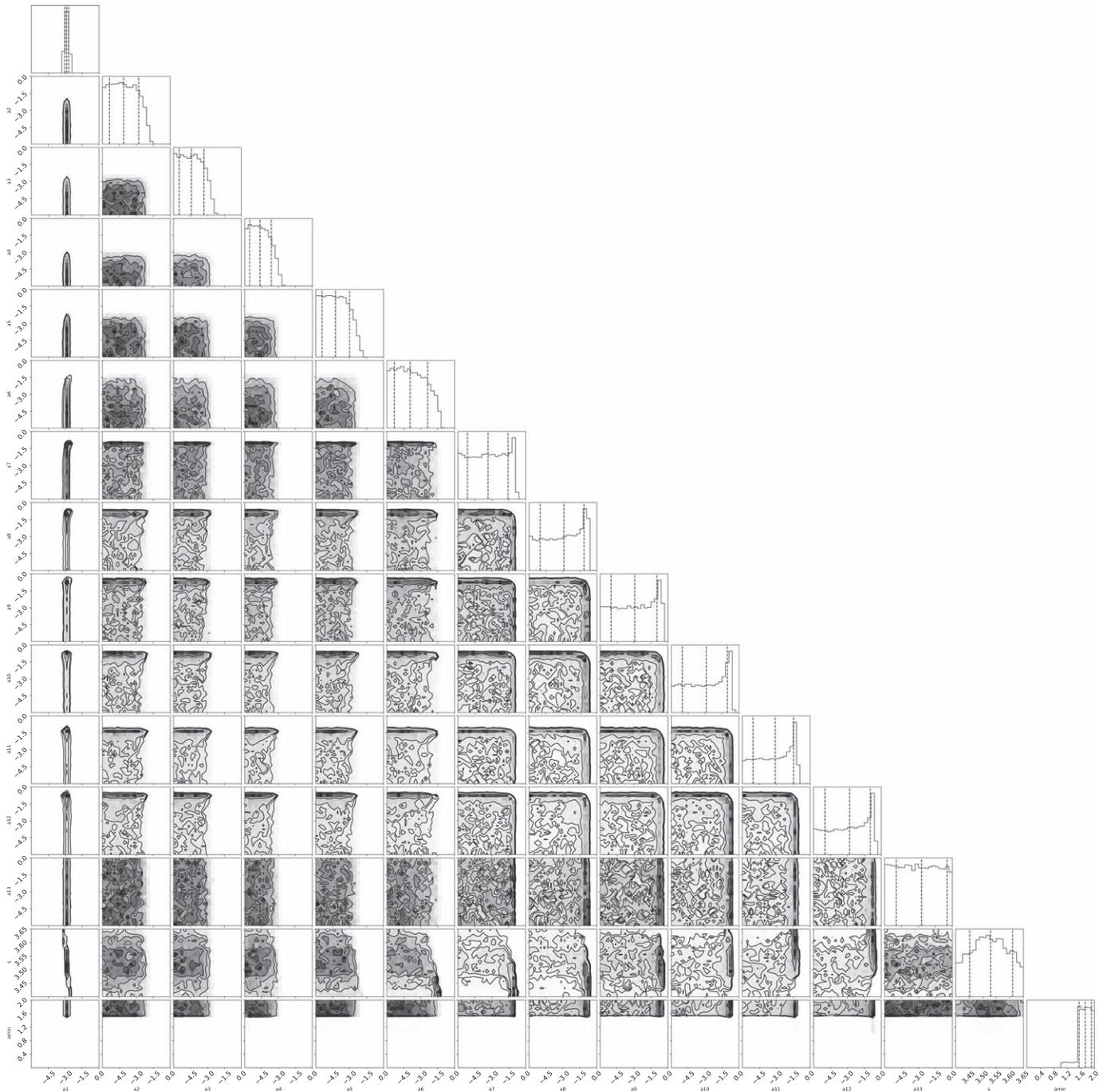

**Figure 19.** Posterior distributions from the MCMC optimized for the full disk broadband SED. The radial extent of each zone is described in Tables 3 and 5, with the latter providing the mass and particle scattering cross sections derived from these scale factors for the best-fit model.

As is the case for the IRS spectra, there is a sharply peaked posterior for the 0–3 au region. However, while the IRS spectrum best-fit model has a dust mass of $\sim 10^{-9}\ M_\oplus$, the SED requires even more dust at this location with the best-fit model having a dust mass of $3 \times 10^{-6}\ M_\oplus$. This could be a consequence of the different particle size distributions used in these two models. The steep slope preferred by the IRS spectrum allows more small dust to be present at that location than the shallower slope preferred by the SED. The agreement between the IRS spectrum and the broadband SED results in this zone provides strong evidence for an Asteroid Belt analog somewhere in the 0–3 au region.

As is the case for the previous SED MCMC run optimized for the outer disk, the best-fit minimum grain size is 2.0 $\mu$m, the largest bin included in our parameter exploration. Consequently, larger minimum grain sizes may have provided a better fit, but require abnormal dust particle compositions, porosities and/or geometries. The particle size distribution slope has a widely peaked posterior distribution centered on $p = 3.5$ while the best-fit model has a negative power-law slope value of 3.65. This agrees well with the result for a collisionally dominated debris disks in a quasi–steady state provided in Gáspár et al. (2012). This slope value is also, unsurprisingly, in the middle ground between the steep slopes preferred by the IRS spectrum and the shallower slopes preferred by the outer disk's SED.

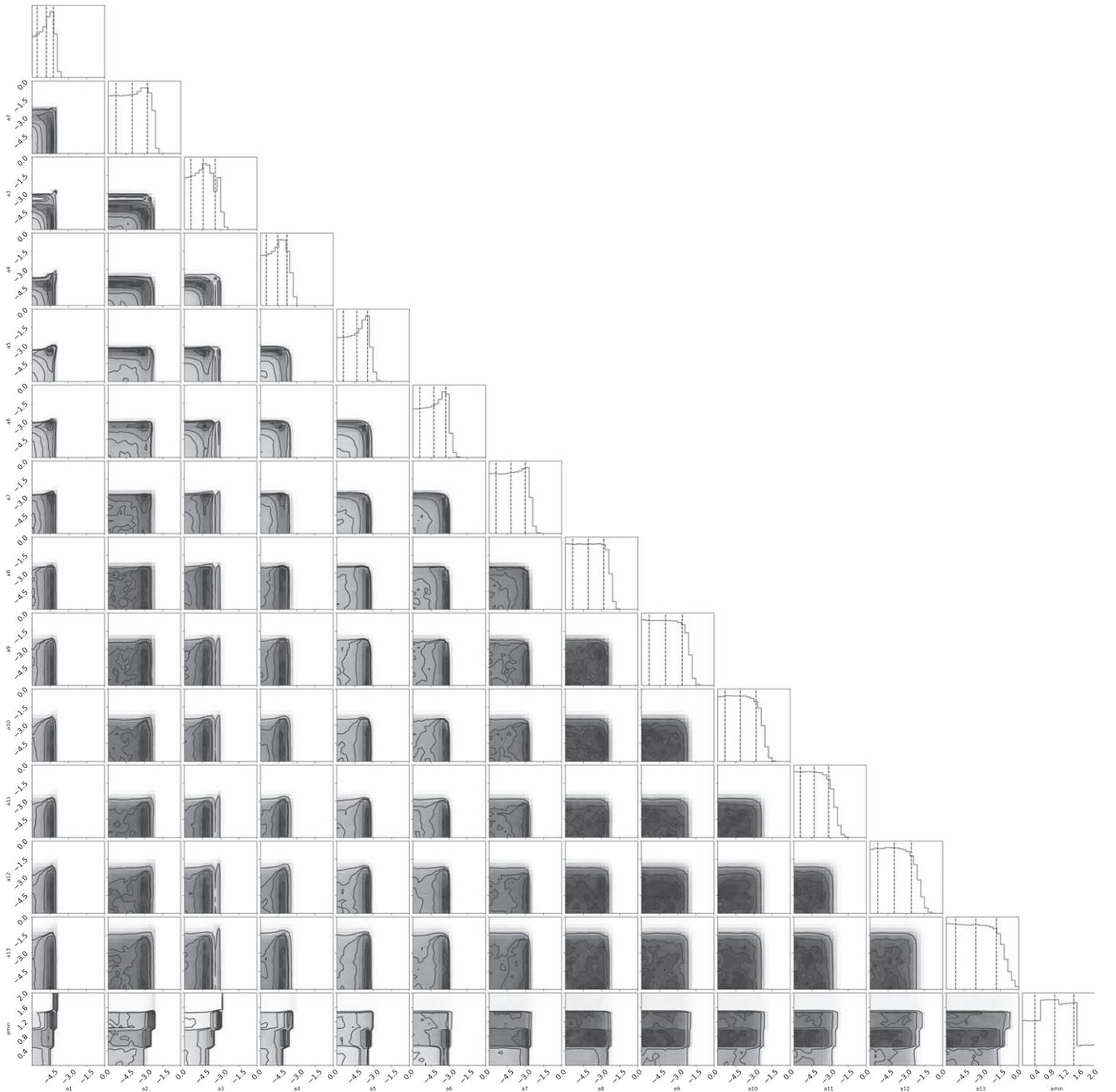

**Figure 20.** Posterior distributions from the MCMC optimized for the Spitzer/IRS spectrum. The radial extent of each zone is described in Tables 3 and 5, with the former providing the mass and particle scattering cross sections derived from these scale factors for the best-fit model. Note that the IRS spectrum is not sensitive to dust in the outer (>37 au) regions of the disk, and the resultant parameter posterior distributions are flat over the allowed range with values for the scale factors of $<10^{-3}$.

For a complete comparison, here we present the full posterior distributions resulting from the MCMC fits to both the Spitzer/IRS spectrum and the broadband photometry. The parameter spaces are degenerate, and these distributions help to visualize correlations between the dust in each radial zone, and the bulk dust properties. Further constraints on the grain albedos from a scattering phase function, or on the intermediate belt locations from planned JWST observations, are required to overcome these limitations.

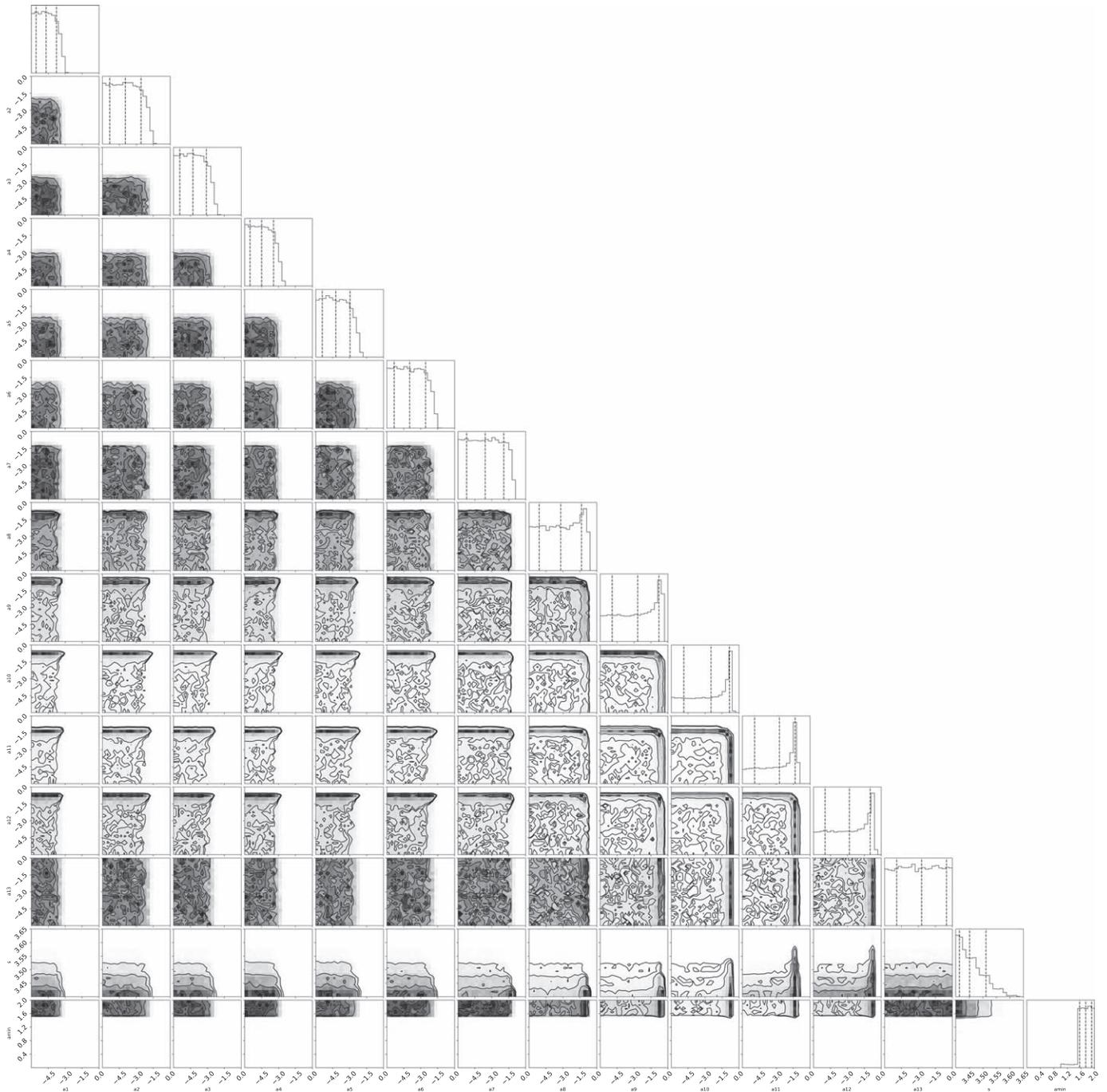

**Figure 21.** Posterior distributions from the MCMC optimized for the broadband SED that has been corrected to only include flux from the outer disk regions ($\gtrsim 10''$). The radial extent of each zone is described in Tables 3 and 5, with the former providing the mass and particle scattering cross sections derived from these scale factors for the best-fit model.

**ORCID iDs**

Schuyler Grace Wolff ● https://orcid.org/0000-0002-9977-8255
András Gáspár ● https://orcid.org/0000-0001-8612-3236
George H. Rieke ● https://orcid.org/0000-0003-2303-6519
Nicholas Ballering ● https://orcid.org/0000-0002-4276-3730
Marie Ygouf ● https://orcid.org/0000-0001-7591-2731